\documentclass[%
 reprint,
 amsmath,amssymb,
 aps,
]{revtex4-1}
\usepackage{float}
\usepackage{graphicx}
\usepackage{dcolumn}
\usepackage{bm}
\usepackage{blindtext}
\usepackage{enumerate}
\usepackage{physics}
\usepackage{tikz}

\def\hext{\hat{H}_{\mathrm{ext}}}
\def\hint{\hat{H}_{\mathrm{int}}}
\def\tautot{\tau_\mathrm{tot}}
\def\phitot{\phi_{{ij}_{\mathrm{tot}}}}

\begin{document}


\title{Dynamic Hamiltonian engineering of 2D rectangular lattices in a one-dimensional ion chain}

\author{Fereshteh Rajabi$^{1}$}
\author{Sainath Motlakunta$^{1}$}
\author{Chung-You Shih$^{1}$}
\author{Nikhil Kotibhaskar$^{1}$}
\author{Qudsia Quraishi$^{3,4}$}
\author{Ashok Ajoy$^{2}$}
\email{ashokaj@berkeley.edu}
\author{Rajibul Islam$^{1}$}
\email{krislam@uwaterloo.ca}

\affiliation{$^{1}$Institute for Quantum Computing and Department of Physics and Astronomy, The University of Waterloo, 200 University Ave. West, Waterloo, Ontario N2L 3G1, Canada
}
\affiliation{$^{2}$Department of Chemistry, University of California Berkeley, and Materials Science Division Lawrence Berkeley National Laboratory, Berkeley, California 94720, USA.}%

\affiliation{$^{3}$Joint Quantum Institute, University of Maryland,
College Park, MD 20742, USA}

\affiliation{$^{4}$Army Research Laboratory, 2800 Powder Mill Rd., Adelphi, MD 20783, USA}

                        
\begin{abstract}

Controlling the interaction graph between spins or qubits in a quantum simulator allows user-controlled tailoring of native interactions to achieve a target Hamiltonian. The flexibility of engineering long-ranged phonon-mediated spin-spin interactions in a trapped ion quantum simulator offers such a possibility. Trapped ions, a leading candidate for simulating computationally hard quantum many-body dynamics, are most readily trapped in a linear 1D chain, limiting their utility for readily simulating higher dimensional spin models. In this work, we introduce a hybrid method of analog-digital simulation for simulating 2D spin models and dynamically changing interactions to achieve a new graph using a linear 1D chain. The method relies on time domain Hamiltonian engineering through a successive application of Stark shift gradient pulses, and wherein the pulse sequence can simply be obtained from a Fourier series decomposition of the target Hamiltonian over the space of lattice couplings. We  focus on engineering 2D rectangular nearest-neighbor spin lattices, demonstrating that the required control parameters scale linearly with ion number. This hybrid approach offers compelling possibilities for the use of 1D chains in the study of Hamiltonian quenches, dynamical phase transitions, and quantum transport in 2D and 3D. We discuss a possible experimental implementation of this approach using real experimental parameters.\\

\begin{description}
\item[PACS numbers]{03.67.Ac, 03.67.Lx, 37.10.Ty}
\end{description}
\end{abstract}
\maketitle
\section{\label{sec:Introduction}INTRODUCTION}

Dynamical evolution of interacting quantum many-body systems are often intractable with classical computation. Controlled studies are best done in quantum simulators \citep{Aspuru2012, Bloch2012, Blatt2012, Cirac2012} wherein the essential many-body dynamics is manifest but resides in an experimentally manageable configuration. Trapped ions~\cite{Blatt2012} are among the most versatile platforms for quantum simulation, especially for simulating quantum spin systems owing to their inherent long-range interactions even when the ions are situated in a 1D topology. Even amongst the gallery of other physical implementations of quantum simulation of spin models, including ultracold bosonic and fermionic atoms in optical lattices \cite{Bloch2005, Bloch2012, Gross2017}, and superconducting circuits \cite{Buluta2009, Houck2012, Devoret2013}, the trapped ion platform offers the versatile and advantageous ability to manipulate individual spin-spin interactions, in principle, arbitrarily \citep{Korenblit2012}. 

Long range spin-spin interactions are exceedingly simple to generate in ion trap quantum simulators, and additionally, can be controlled in their range, magnitude, and sign \cite{Molmer1999, Deng2005, Kim2009, Kim2010, Britton2012, Islam2013, Richerme2014, Jurcevic2014, Bohnet2016}. Leveraging phonon modes to build the inter-spin interactions makes a trapped ion system fully-connected and so \textit{inherently} higher dimensional, allowing potentially the ability to probe a rich variety of physical phenomena, such as quantum transport and localization, topological insulators \cite{Qi2011}, the Haldane model \cite{Haldane2008}, as well as in topological quantum computation following the Kitaev honeycomb model \cite{Kitaev2003, Schmied2011} and can be advantageous for quantum computing \cite{Linke2017}.

However, despite a few notable experiments and proposals \cite{ Sawyer2012, Britton2012, Yoshimura2015, Bohnet2016, Richerme2016, Li2017}, most quantum simulations have been limited to one-dimensional (1D) chain of ions due to the constraints of radio-frequency ion traps \cite{Wineland1998}. While experimental efforts to broaden the number of ion traps with higher dimensional ion arrays are underway, significant experimental simplification is offered by leveraging existing 1D ion chains, especially considering remarkable progress, where $N>$100 ions have been trapped in a linear geometry~\cite{Zhang2017,Pagano2018}, and prospects for still larger system sizes looking optimistic in the future. Additionally, existing simulators can be experimentally resource-intensive, operating on either analog \cite{Friedenauer2008, Kim2010,Gerritsma2010, Gerritsma2011, Islam2011, Britton2012, Islam2013, Richerme2014, Jurcevic2014, Senko2015, Bohnet2016, Jurcevic2017, Zhang2017} or digital \cite{Lanyon2011, Barreiro2011, Linke2017, Hempel2018} quantum simulation protocols. Whereas an analog approach requires fine tuning of several experimental parameters for careful frequency, amplitude and phase modulation of optical beams, leading to control parameters that often scale quadratically with ion number; a digital approach requires individual beams for each ion and consequently stringent control over optical beam paths of the same number as there are ions. 

\begin{center}
 \begin{figure*}[ht]
        \includegraphics[angle=270, width = 0.8\textwidth]{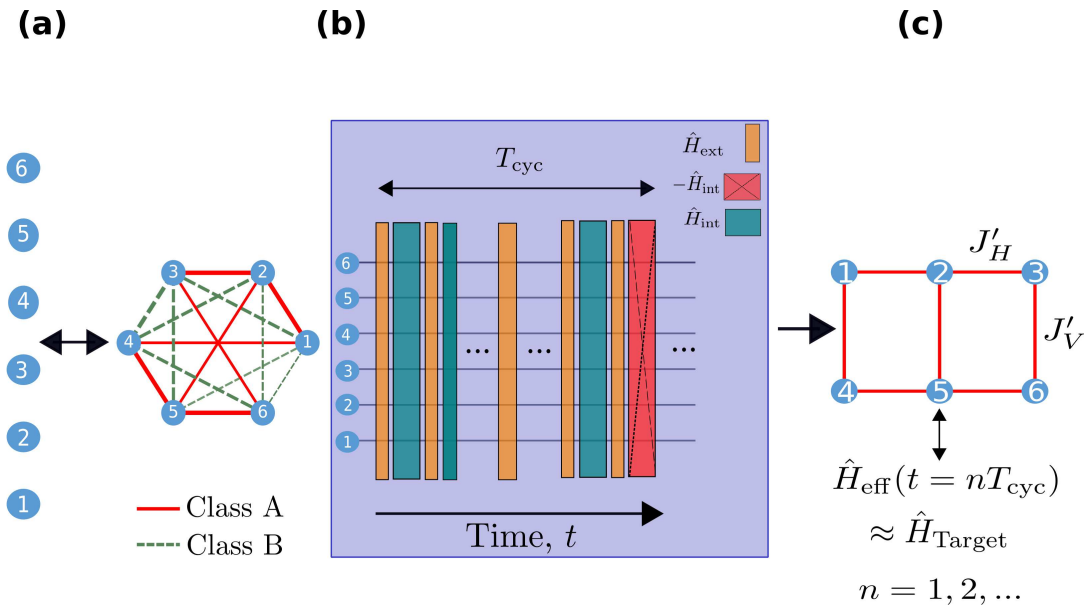}
        \caption{Schematic of a hybrid analog-digital quantum simulation of a 2D rectangular lattice via dynamic Hamiltonian engineering. a) A 1D chain of $N=6$ ions acts like a fully connected network of spins as a result of long range phonon-mediated spin-spin interactions. Here the thickness of bonds represent the strength of different couplings. The interactions in the ion network can be categorized into two classes, A (shown in solid red) and B (shown in dashed green). Only class A bonds are present in the target Hamiltonian. b) The interaction network can be modified by subjecting the system to a periodic sequence of free evolution under the native Hamiltonian $\pm\hat{H}_{\mathrm{int}}$ and single qubit (phase) gates $\hat{H}_{\mathrm{ext}}$ acting simultaneously on all qubits. c) The average Hamiltonian of the system, $\hat{H}_\mathrm{eff}$ resembles the target Hamiltonian $\hat{H}_\mathrm{Target}$, here a $2\times 3$ rectangular lattice, at discrete time steps, $t =n T_{\mathrm{cyc}}$ ($n=1,2,\cdots$). The horizontal and the vertical couplings of the target Hamiltonian, $J'_H$ and $J'_V$ respectively, can be dynamically changed in a simulation. A square lattice is obtained when $J'_H=J'_V$.}
        \label{fig:DigitalAnalog}
 \end{figure*}
\end{center}

In this work, we propose a hybrid method of analog-digital quantum simulation that can allow the \textit{dynamic} engineering of a fully-connected 1D ion chain to, in principle, an arbitary 2D lattices (see Figure~\ref{fig:DigitalAnalog}).  When the target lattice contains certain symmetries, for instance in the case of engineering a 2D rectangular lattice, the quantum control required ($\mathcal{O}(N)$) scales exceedingly favorably compared to other methods.  The method relies on a repeated and stroboscopic application~\cite{Warren1979, Baum1985, Ajoy2013} of the full interaction Hamiltonians $\hint$ and -$\hint$, and laser driven Stark shift gradients, allowing the \textit{time-domain} engineering of the interaction graph. In an analogy to holography, the exact Hamiltonian engineering is efficent in the Fourier domain of couplings in the interaction graph, allowing a powerful means to engineer the target Hamiltonian while exploiting its inherent symmetries. Indeed, as we shall demonstrate, the time-sequence of Hamiltonians to be applied, can be simply read-off from a Fourier series expansion of the target Hamiltonian graph in an appropriate encoding space.  Most importantly, since the engineered Hamiltonians can be dynamically modified, this opens several possibilities for studying quantum transport, dynamical phase transitions under a Hamiltonian quench \citep{Heyl2013, Vosk2014, Jurcevic2017, Zhang2017}, and thermalization \citep{Rigol2008} and many-body localization \citep{Schreiber2015, Smith2016, Bordia2016, Choi2016, Luschen2017} in high dimensions.

Figure~\ref{fig:DigitalAnalog} shows a schematic of the Hamiltonian engineering scheme. It works by removing (\textit{decoupling}) interactions (forthwith ``class B'' interactions) that are absent in the target Hamiltonian graph, while appropriately weighting (\textit{engineering}) the other interactions (``class A''), all by the \textit{global} manipulation of all spins in the linear ion chain. Thus, the experimental implementation is considerably simpler than a fully digital simulation model, which requires individual two qubit gates on the ion chain.  Practically, the global spin-spin interactions ($\pm \hat{H}_\mathrm{int}$) are realized by laser driven M{\o}lmer-S{\o}rensen couplings \cite{Molmer1999} and the single qubit phase gates (by $\hat{H}_\mathrm{ext}$) are realized by imprinting light shift (AC Stark shift) in the qubit frequency by an additional laser beam with an intensity gradient. The sign of the internal Hamiltonian ($\pm \hat{H}_\mathrm{int}$) can be flipped by changing the frequencies of global laser beams \cite{Bohnet2016}. The scheme can be extended to other 2D lattice geometries, 3D lattices, and can potentially be adapted to other systems with long-range interactions and control over individual spins. Our approach therefore offers both a simplification of control parameters and favorable scaling with ion number, and offers compelling possibilities for exploiting the remarkable versatility of long-range coupled linear chain of ions for the generation of exotic engineered Hamiltonians.

\section{\label{sec:LongRangeInteractions}Long range spin-spin interactions in trapped ions}

Internal electronic or hyperfine states of trapped ions act as pristine spin-1/2 objects, with coherences extending to many minutes \cite{Wang2017}. These long coherence times are useful for implementation of control pulse sequences.  In a typical radio-frequency Paul trap, multiple ions can be laser-cooled to a linear chain configuration in an anisotropic confinement potential. The Coulomb repulsion between ions results in collective phonon or vibrational normal modes. Off-resonant optical dipole forces exerted by laser beams or global microwave radiation with a strong spatial field gradient \cite{Ospelkaus2011, Cohen2015} can induce spin-phonon couplings, resulting in phonon-mediated spin-spin interactions \cite{Cirac1995, Molmer1999, Milburn2000, Leibfried2003, Porras2004, Kim2009}. For example, a long range flip-flop or XY Hamiltonian, 
\begin{equation}\label{Equ:H_XY}
\hat{H}_{\mathrm{int}} = \sum_{i<j} J_{ij} \left(\hat{S}^{+}_{i}\hat{S}^{-}_{j}+ \hat{S}^{-}_{i}\hat{S}^{+}_{j}\right) 
\end{equation} has been engineered in recent experiments \citep{Richerme2014, Jurcevic2014}, where $\hat{S}^{\pm} = \hat{S}_x \pm i \hat{S}_y$ are the raising and lowering spin operators. The interactions in Eq. \ref{Equ:H_XY} can be tuned according to a power law, 
\begin{equation}\label{Equ:Jij}
J_{\mathrm{ij}} \approx \frac{J_0}{\mid i - j \mid ^{\alpha}}.
\end{equation}
Here, $0<\alpha<3$ sets the range of interactions \citep{Porras2004,Deng2005} and $J_0$ is the nearest neighbor coupling strength. Without loss of generality, we assume $J_0>0$. More generally, with full control over individual spin-phonon couplings, it has been shown that $J_{\mathrm{ij}}$ can be arbitrarily programmed \cite{Korenblit2012}. However, controlling all the spin-phonon couplings is a challenging experimental task and may not be scalable with current technologies. As a special case of Eq. (\ref{Equ:Jij}), the range of interactions can be quasi-infinite range, $\alpha\approx 0$ when the field driving coherent spin-phonon coupling is tuned close to the center of mass (COM) phonon mode. Further, by changing frequency of laser beams, the sign of interactions \cite{Kim2009, Bohnet2016} can be flipped (see Appendix~\ref{sec:Hint_to_minusHint}).

For a fully connected system of $N$ spins, there are $\binom{N}{2} = \frac{N(N-1)}{2}$ couplings. To engineer the target interaction graph, we categorize these couplings into two classes, which we name class A and class B (see Fig.~\ref{fig:DigitalAnalog}a). The couplings in class A are present in the target graph, while those in class B need to be removed. Further, the couplings in class A may need additional scaling depending on the target interaction graph. For example, in Fig.~\ref{fig:DigitalAnalog}c, the vertical nearest neighbor bonds $J'_V$ in the target square lattice are obtained from the third neighbors ($J_{i,i+3}$) in the original 1D chain, while the horizontal nearest neighbor bonds $J'_H$ are obtained from the nearest neighbors ($J_{i,i+1}$ except $J_{34}$) in the 1D chain. In a square lattice, we require $J'_H=J'_V$ and hence our engineering protocol must reduce the strength of the nearest neighbor couplings (except $J_{34}$) in the 1D chain to match the third neighbor couplings if $\alpha > 0$ in Eq.~\ref{Equ:Jij}. This is because we cannot enhance the strength of a coupling or engineer one that does not already exist in the original network. The protocol for Hamiltonian engineering is described in detail in section \ref{sec:Method}.
 
\section{\label{sec:Method}METHOD}
\subsection{\label{sec:HintSignFlip}Cancellation of couplings by reversing unitary time evolution}
\begin{center}
   \begin{figure*}
    \includegraphics[width = 0.85\textwidth]{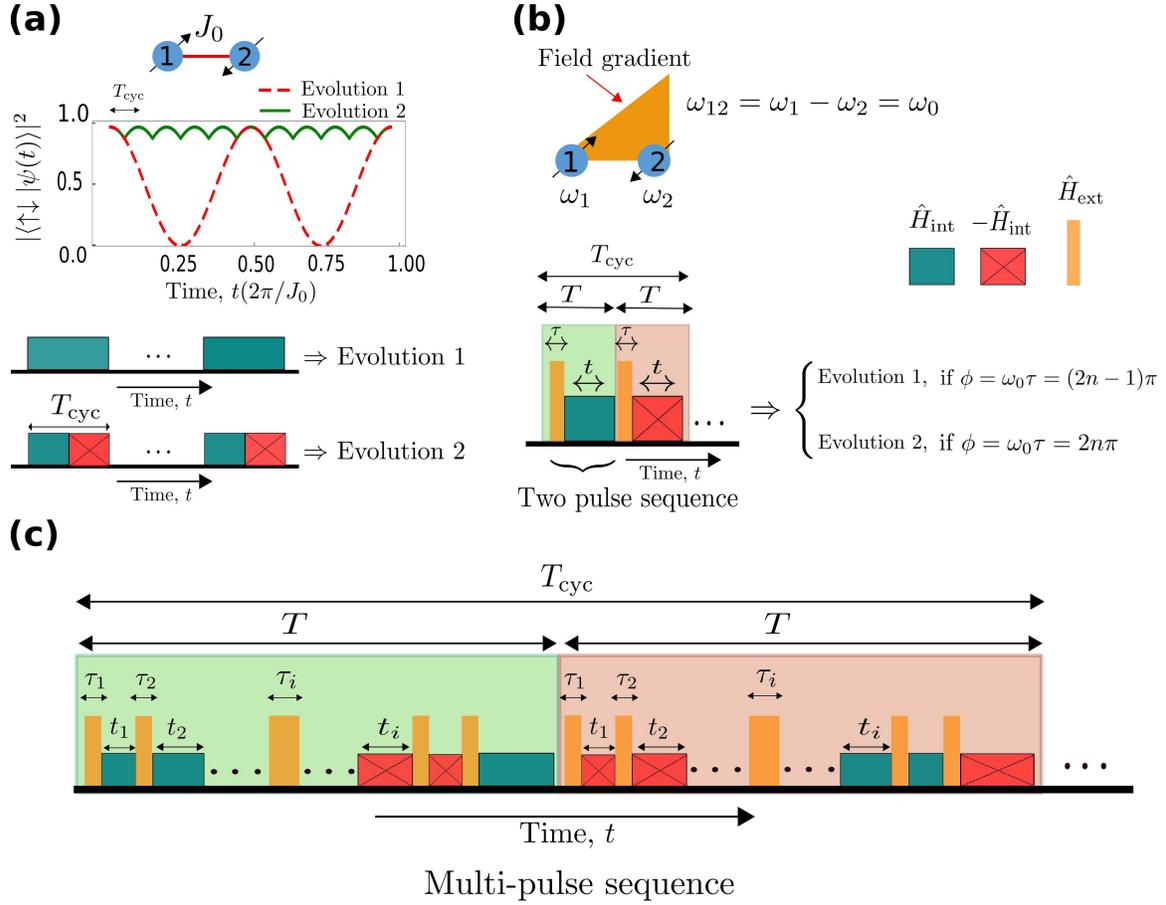}
        \caption{Protocol for canceling and scaling interactions. a) Continuous time evolution of $N=2$ spins under a flip-flop Hamiltonian, $\hint$ (Eq.~\ref{Equ:H_XY}) is shown by the dashed red line (labeled as Evolution 1), where the spins are initialized in $\mid\uparrow \downarrow\rangle$. If the Hamiltonian is switched from $\hint$ to $-\hint$ halfway through a time cycle of duration $T_\mathrm{cyc}$, the unitary time evolution (solid green line labeled as Evolution 2) is reversed and the spins return to the initial state at $t=nT_\mathrm{cyc}$ ($n=1,2,\cdots$). Thus, the interactions are effectively canceled when the system is viewed at discrete time steps, $t=nT_\mathrm{cyc}$. b) Evolution 1 and Evolution 2 are controllably reproduced by the same pulse sequence, if we introduce an external gradient field (Eq.~\ref{Equ:HStark}). In this scheme, one cycle in the time evolution consists of two subsequent blocks. The first block consists of $\hat{H}_{\mathrm{ext}}$ for time $\tau$ followed by $\hat{H}_{\mathrm{int}}$ for time $t$ and the second block consists of $\hat{H}_{\mathrm{ext}}$ for time $\tau$ followed by $-\hat{H}_{\mathrm{int}}$ for time $t$. The $\hat{H}_{\mathrm{ext}}$ pulse assigns a phase tag $\phi = \omega_0\tau$ to the coupling between the ions, where $\omega_0=\omega_2-\omega_1$. If the assigned phase $\phi =(2n-1)\pi$, $n$ being an integer number, Evolution 1 is reproduced and the interaction is retained. If $\phi = 2n\pi$, Evolution 2 is reproduced and the interaction is effectively canceled. c) To rescale the interactions, the two-pulse sequence is replaced by a multi-pulse sequence. Each block now consists of alternate blocks of $\hat{H}_{\mathrm{ext}}$ and $\pm\hat{H}_{\mathrm{int}}$ for different durations. The duration of the multi-pulse sequence cycle $T_{\mathrm{cyc}}$ is $2T$, where $T$ is the duration of each block. In alternate blocks $\pm\hat{H}_{\mathrm{int}}$ are switched with $\mp\hat{H}_{\mathrm{int}}$ \cite{Baum1985}.}       \label{fig:HintSignFlip_evolution2}
    \end{figure*}
 \end{center}

One way to effectively cancel spin-spin interactions is subjecting the system to periodic time evolutions alternating under $\hat{H}_{\mathrm{int}}$ and $-\hat{H}_{\mathrm{int}}$ for equal durations. The unitary time evolution under $\hint$ is reversed under $-\hint$, returning to the initial state. In Fig.~\ref{fig:HintSignFlip_evolution2}a, we illustrate this for $N=2$ spins. The spins, when initialized in the product state $\mid \uparrow \downarrow\rangle$ will undergo coherent oscillation between $\mid \uparrow \downarrow\rangle$ and $\mid \downarrow \uparrow\rangle$ under the flip-flop Hamiltonian, Eq.~(\ref{Equ:H_XY}). Here, $\mid \uparrow \rangle$ and $\mid \downarrow \rangle$ are the eigenstates of $\hat{S}_z$. If the Hamiltonian is instead switched alternately between $\hat{H}_{\mathrm{int}}$ and $-\hat{H}_{\mathrm{int}}$ the spins go back to their initial state after each cycle of duration $T_\mathrm{cyc}$. Thus, when observed discretely at time $t=nT_\mathrm{cyc}$ ($n=0,1,2,\cdots$), the spins appear to be non-interacting. However, we must need additional ingredients in this protocol as we want to protect interactions in class A. Our protocol must distinguish between couplings that we want to cancel by reversing unitary time evolution and couplings that we want to rescale. This is done by imprinting separate `phase tags' between these classes of couplings by a spatial field gradient applied simultaneously on all the spins \cite{Ajoy2013}. The Hamiltonian from this external field is,
\begin{equation}\label{Equ:HStark}
    \hat{H}_{\mathrm{ext}} = \sum_{i=1}^{N} \omega_{i} \hat{S}_{z_{i}},
\end{equation}
where $\{\omega_{i}\}$ depends on the nature of the field gradient. $\hat{H}_{\mathrm{ext}}$ can be engineered in experiments, e.g., by a laser beam with spatially inhomogeneous intensity pattern imprinting AC Stark shifts on the spins. 

If the external Hamiltonian, Eq. (\ref{Equ:HStark}), is applied for a duration $\tau$, the Hamiltonian in Eq. (\ref{Equ:H_XY}) is transformed to $\hat{\tilde{H}}_{\mathrm{int}} = e^{i\hat{H}_{\mathrm {ext}} \tau} \hat{H}_{\mathrm{int}}e^{-i\hat{H}_{\mathrm {ext}}} = \sum_{i<j} J_{ij} \hat{S}^{+}_{i}\hat{S}^{-}_{j} e^{i\omega_{ij}\tau} + h.c.$ in the rotating frame of the external field, where $\omega_{ij} = \omega_{i}-\omega_{j}$. Thus the phase tags $\phi_{ij} =\omega_{ij} \tau$ appear in $\hat{\tilde{H}}_{\mathrm{int}}$. 
We want to choose $\{\omega_{i}\}$ in Eq.~(\ref{Equ:HStark}) such that couplings in class A accumulate a phase $\phi^{\mathrm{(A)}}_{{ij}} = (2n - 1)\pi$ and couplings in class B accumulate a phase $\phi^{\mathrm{(B)}}_{{ij}} = 2n\pi$, with $n$ being an integer. Thus, the interactions in class A evolve under  $\hat{\tilde{H}}_{\mathrm{int}} = e^{-i(2n-1)\pi} \hat{H}_{\mathrm{int}} = -\hat{H}_{\mathrm{int}}$, whereas interactions in class B evolve under $\hat{\tilde{H}}_{\mathrm{int}} = e^{-i2n\pi} \hat{H}_{\mathrm{int}} = \hat{H}_{\mathrm{int}}$. The external field gradient and switching between $\hat{H}_{\mathrm{int}}$ and $-\hat{H}_{\mathrm{int}}$ can be combined in the time evolution cycle to preserve interactions in class A while canceling interactions in class B. Fig. \ref{fig:HintSignFlip_evolution2}b demonstrates the basic principle for $N=2$ spins. One cycle in the time evolution consists of two subsequent blocks of pulses. The first block consists of $\hat{H}_{\mathrm{ext}}$ for time $\tau$ followed by $\hat{H}_{\mathrm{int}}$ for time $t$, and the second block consists of $\hat{H}_{\mathrm{ext}}$ for time $\tau$ followed by $-\hat{H}_{\mathrm{int}}$ for time $t$. Setting $\phi_{ij} = (2n-1)\pi$ preserves the interaction between the $(i,j)$ pair of spins (class A), while $\phi_{ij} = 2n\pi$ cancels the interaction (class B).
\subsection{\label{subsec:multiplepulse} Scaling of couplings by a multi-pulse sequence}
To rescale the couplings $\{J_{ij}\}$, the simple two pulse sequence can be replaced by a multi-pulse sequence acting as a filtering function, as schematically shown in Fig.~\ref{fig:HintSignFlip_evolution2}c. Each block now consists of alternate pulses of $\hat{H}_{\mathrm{ext}}$ for time $\tau_i$ and $\pm\hat{H}_{\mathrm{int}}$ for time $t_i$. The total phase accumulated by the $(i,j)$ spin-pair from $\hat H_{\mathrm{ext}}$ must satisfy $\phi_{{ij}_{\mathrm{tot}}} = \omega_{ij}\tau_{\mathrm{tot}}= (2n-1) \pi$ for it to not cancel under the reversal of unitary evolution by $-\hint$. Here, $\tau_{\mathrm{tot}}=\sum_{k=1}^l{\tau_k}$, $l$ being the number of pulses of $\hext$ in each block. The unitary evolution of the system at $t = T$, the end of a multi-pulse block, is given by
\begin{equation}
    \hat{U}(T) = e^{-i\hat{H}_{{\mathrm {s}}}t_l}e^{-i\hat{H}_{{\mathrm {ext}}} \tau_l}\cdots e^{-i\hat{H}_{{\mathrm {ext}}} \tau_2} e^{-i\hat{H}_{{\mathrm {s}}}t_1} e^{-i\hat{H}_{{\mathrm {ext}}} \tau_1}
\end{equation}
where $\hat{H}_{\mathrm{s}} = \pm\hat{H}_{\mathrm{int}}$. This equation can be rewritten in the following form 
\begin{equation}\label{eq:togglingframe}
    \hat{U}(T) = e^{-i\hext\tau_{\mathrm{tot}}} e^{-i\hat{\tilde{H}}_{{\mathrm {s}}_l} t_l}\cdots e^{-i\hat{\tilde{H}}_{{\mathrm {s}}_2} t_2} e^{-i\hat{\tilde{H}}_{{\mathrm {s}}_1} t_1},
\end{equation}
where $\hat{\tilde{H}}_{{\mathrm {s}}_k} = e^{i\hat{H}_{\mathrm {ext}} \sum_{j=1}^k \tau_j} \hat{H}_{\mathrm{s}}e^{-i\hat{H}_{\mathrm {ext}}\sum_{j=1}^k \tau_j}$ is the internal Hamiltonian in the rotating frame defined by successive frame transformation induced by the external gradient field. Eq. \ref{eq:togglingframe} is derived using $\hat{I} = e^{-i\hat{H}_{\mathrm {ext}}\tau_j}e^{i\hat{H}_{\mathrm {ext}} \tau_j}$. Upon applying the average Hamiltonian theory \cite{ Haeberlen1968,Warren1979, Maricq1982}, an effective Hamiltonian for the full block can be defined as
\begin{align}
    \hat{H}_{\mathrm{eff}} (T) &\approx \hat{H}^{(0)}_{\mathrm {avg}} (T) \nonumber \\
    & = \frac{1}{T}\left(\int_{0}^{t_1} \hat{\tilde{H}}_{{\mathrm {s}}_1}(t') dt'+\cdots+\int_{T-t_l}^{T} \hat{\tilde{H}}_{{\mathrm {s}}_l}(t') dt'\right) \nonumber \\
    & = \frac{1}{T}\left( t_1 \hat{\tilde{H}}_{{\mathrm {s}}_1} + t_2 \hat{\tilde{H}}_{{\mathrm {s}}_2}\cdots +t_l\hat{\tilde{H}}_{{\mathrm {s}}_l}\right)\label{eq:Heff_multipleBlock2}
\end{align}
so that 
\begin{equation}\label{eq:evolutionHeff}
    \hat{U}(T) = e^{-i\hext\tautot} e^{-i\hat{H}_{\mathrm{eff}} T}.
\end{equation}
Here we have assumed that $\sum_{j=1}^l t_j\approx T$, which is a valid approximation if $\tautot\ll T$ and can be realized with the external gradient field much stronger than the spin-spin interactions. Eq.~(\ref{eq:evolutionHeff}) implies that the the multi-pulse block can be thought of as a single pulse of $\hext$ applied for a duration of $\tautot$ followed by a pulse of $\hat{H}_{\mathrm{eff}}$ for a duration of $T$. Note that, in order for the average Hamiltonian formalism (Eq.~(\ref{eq:Heff_multipleBlock2})) to be valid, $J_0T\ll 1$.

Eq.~(\ref{eq:Heff_multipleBlock2}) can be written in a more explicit form for the coupling $J_{ij}'$ of $\hat{H}_{\mathrm{eff}}$ in terms of the $J_{ij}$ of $\hint$.
\begin{align*}
    J_{ij}'(T) = \frac{J_{ij}}{T}[& \pm t_1 e^{i\phi_{{ij}_1}} \pm t_2 e^{i(\phi_{{ij}_1} + \phi_{{ij}_2})} \pm \cdots\\
    & \pm t_le^{i(\sum_{k=1}^l\phi_{{ij}_k})}] \\
    & \equiv \beta_{ij} J_{ij}
\end{align*}
where $\phi_{{ij}_k} = \omega_{ij}\tau_{k}$ and $\sum_{k=1}^l\phi_{{ij}_k} = \phi_{{ij}_{\mathrm{tot}}}$, the total phase accumulated in the multi-pulse block. The real-valued scaling parameter $\beta_{ij}$ is explicitly given by,
\begin{equation}\label{eq:betaij}
    \beta_{ij}=\frac{1}{T}\left[ \pm t_1 e^{i\phi_{{ij}_1}} \pm t_2 e^{i(i\phi_{{ij}_1}+\phi_{{ij}_2})} \cdots \pm t_l e^{i\phi_{{ij}_{\mathrm{tot}}}}\right],
\end{equation}
and can be engineered by choosing $\{t_k,\tau_k\}$ (for $k=1,2,\cdots,l$) for a chosen $\{\omega_i\}$.
If the multi-pulse block is repeated while all $\pm\hat{H}_{\mathrm{int}}$ are replaced with $\mp\hat{H}_{\mathrm{int}}$, the effective interactions at the end of the cycle, $t = T_\mathrm{cyc}=2T$, becomes \cite{Ajoy2013}
\begin{equation}\label{eq:Heff_twoBond}
    J_{ij}'(T_\mathrm{cyc}) = \frac{J_{ij}}{2}[\beta_{ij} (1 - e^{i\phi_{{ij}_{\mathrm{tot}}}})].
\end{equation}
Thus, if $\phi_{{ij}_{\mathrm{tot}}} = (2n-1)\pi$, 
\begin{equation}\label{eq:Heff_twoBond2}
    J_{ij}'(T_\mathrm{cyc}) = \beta_{ij} J_{ij}, 
\end{equation}
and the couplings are rescaled.
While, if $\phi_{{ij}_{\mathrm{tot}}} = 2n\pi$, the couplings vanish automatically regardless of $\beta_{ij}$,
\begin{equation}\label{eq:Heff_twoBond3}
    J_{ij}'(T_\mathrm{cyc}) = 0. 
\end{equation}

Eqs.~(\ref{eq:Heff_twoBond})-(\ref{eq:Heff_twoBond3}) indicate that we only need to consider those pairs with $\phi_{{ij}_{\mathrm{tot}}} = (2n-1)\pi$ in designing the scaling parameter $\beta_{ij}$. As we will show in Section \ref{subsec:Fourier_MultipulseSequence}, the pulse sequence to engineer the target scaling factors (Eq.~(\ref{eq:betaij})) can be constructed from a Fourier series expansion of the target Hamiltonian in the space of the interactions.  
\subsection{\label{subsec:FieldGradient} Labeling and field gradient scheme}

\begin{center}
   \begin{figure}
       \includegraphics[width = 0.97\columnwidth]{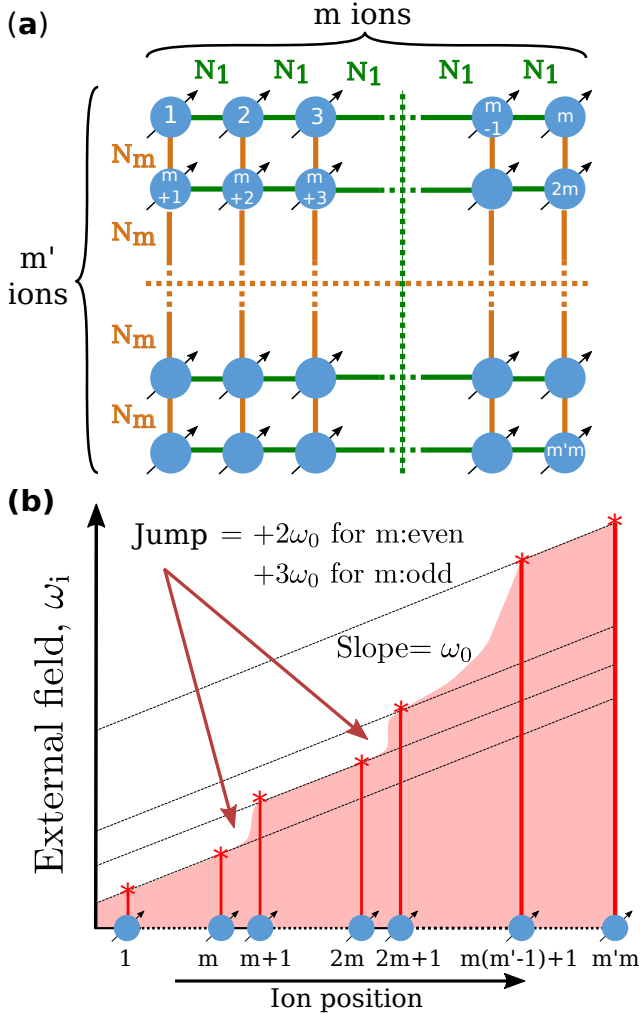}
        \caption{Labeling and field gradient scheme for $m' \times m$ rectangular lattices ($m'< m$). a) We employ a labeling scheme where $N_1$ (the nearest neighbor in the 1D chain) and $N_m$ (the $m^\mathrm{th}$ neighbor in the 1D chain) form, respectively, the horizontal and vertical bonds of the target lattice. b) A semi-linear external field gradient ($\{\omega_i\}$ in Eq.~(\ref{Equ:HStark})) provides the necessary conditions, described in the text, for efficiently canceling and scaling interactions to achieve the target graph. In this profile, $\omega_i$ increases linearly with a constant slope $\omega_0$ with some added jumps of $+2\omega_0$ (for even $m$) and $+3\omega_0$ (for odd $m$). These jumps are designed between the $km^{\mathrm{th}}$ and $(km+1)^{\mathrm{th}}$ ions, such that the torroidal couplings $J_{km,km+1}$ can be canceled ($k = 1, 2,\dots, m'-1$). See Fig.~\ref{fig:9ion_labelling_gradient} in Appendix \ref{sec:9-ions} for an example.}
        \label{fig:labeling/gradient_scheme}
    \end{figure}
 \end{center}

To map the interactions $J_{ij}$ in the 1D spin-chain onto the target 2D rectangular lattice, we categorize them into classes $N_d$ according to the distance $d=|j-i|$ between the spins, with $N_1$ denoting the nearest neighbor couplings, $N_2$ denoting the next nearest-neighbor couplings and so on. Here, we ignore inhomogeneities due to the finite size effect, which we discuss in Section \ref{subsec:Discussion}. As seen in Fig. \ref{fig:labeling/gradient_scheme}a, the $N_1$ and $N_m$ couplings form the horizontal and vertical bonds of an $m'\times m$ rectangular lattice. That is class $A = \{N_1, N_m\}$, with the exception of couplings $J_{km,km+1}$, $k = 1,...,m'-1$, that form a toroidal linkage between the edges of the rectangular lattice and must be excluded from class A. Thus, we need an external gradient field $\{\omega_i\}$ that assigns,
\begin{enumerate}[1)]
    \item $\phitot = (2n-1)\pi$ to all couplings in class $A = \{N_1, N_m\}$. The integer $n$ does not have to be the same for all of the couplings. However, the problem of designing the filtering function simplifies with a small number of different $n$'s,
    \item distinct phase tags to $J_{km,km+1}$ from other couplings in class A,
    \item $\phitot = 2n\pi$ to as many couplings as possible in class B,
    \item $\phitot = (2n-1)\pi$ to couplings in class B for which it is not possible to assign $\phitot = 2n\pi$ after satisfying the constraints for couplings in class A. These couplings have to be rescaled to zero ($\beta_{ij}=0$) by the Fourier filtering function.
\end{enumerate}

A semi-linear field gradient as shown in Fig. \ref{fig:labeling/gradient_scheme}b satisfies the conditions above. We choose the external field profile to increase linearly with a constant slope $\omega_0$ with added jumps of $2\omega_0$ (for even $m$) or $3\omega_0$ (for odd $m$)  between the $km^\textrm{th}$ and $(km+1)^\textrm{th}$ spins to address the toroidal linkages $J_{km,km+1}$. Hence,
\begin{center}
\begin{tabular}{lll}
    $\omega_{i,i+1}$ & $=$ & $\omega_0$ for $N_1$ except $J_{km,km+1}$, \\
    $\omega_{i,i+m}$ & $=$ & $(m+2)\omega_0$ for $N_m$ when m is odd, \\
    $\omega_{i,i+m}$ & $=$ & $(m+1)\omega_0$ for $N_m$ when m is even.
\end{tabular}
\end{center}

So, $\omega_0 \tautot = \pi$ satisfies the conditions above for all couplings within class A and some of the couplings in class B. 
 \subsection{\label{subsec:Fourier_MultipulseSequence}Fourier filtering of interactions}
The interactions that were not automatically be canceled by our chosen field gradient can be rescaled to their target value by designing a Fourier filter.
The $N_1$ couplings in class A have to be rescaled to match the $N_m$ couplings for $\alpha\ne 0$ in Eq.~(\ref{Equ:Jij}). In addition, the couplings in class B for which $\phitot=(2n-1)\pi$ have to be rescaled to zero. The scaling is performed through a Fourier filter function $F(\phi)$ that produces the desired scaling parameter $\beta_{ij}$ in Eq.~(\ref{eq:Heff_twoBond2}). The filter function $F(\phi)$ is defined as, 
 \begin{equation}\label{Equ:Fourier}
    F(\phi) = a_0 + \sum_{i^{\prime}=1}^i a_{i^{\prime}} \cos(i^{\prime} W \phi). 
 \end{equation}
 The filter function $F(\phi)$ should satisfy $F(\phi_{{ij}_{\mathrm{tot}}})=\beta_{ij}$ for all couplings with a phase $\phitot=(2n-1)\pi$. 
 \begin{center}
   \begin{figure}
        \includegraphics[width=\columnwidth]{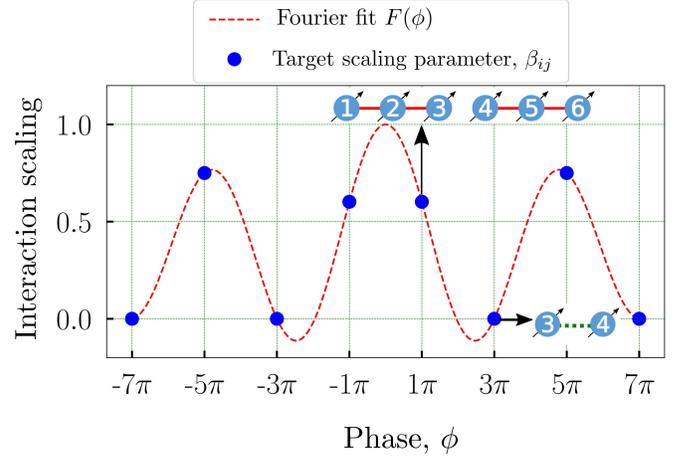}
        \caption{Constructing the Fourier filtering function. For a $2\times 3$ square lattice, the filtering function, $F(\phi)$ in Eq.~(\ref{Equ:Fourier}) is fit to the Fourier series target scaling parameters, $\beta_{ij}$ in Eq.~(\ref{eq:Heff_twoBond2}). The field gradient in Eq.~(\ref{Equ:HStark}) creates a phase of $\phi=\phi_{ij}=\omega_{ij}\tautot$ for all the couplings $J_{ij}$ that need to be rescaled by the same factor $\beta_{ij}$. For this target lattice and the field gradient of Fig. \ref{fig:labeling/gradient_scheme}(b), there are four relevant phases, $\pi$, $3\pi$, $5\pi$, and $7\pi$ given the constraints set by Eq.~(\ref{eq:Heff_twoBond2}) and Eq.~(\ref{eq:Heff_twoBond3}). The points with negative phases are included to satisfy the symmetry condition of Eq.~(\ref{Equ:Fourier}). All the nearest neighbors, $N_1$, except $J_{34}$, acquire a phase of $\phi=\pi$. They are scaled by a factor of $1/3^\alpha=0.8$ with respect to $N_3$ to attain a square lattice geometry, if the interactions in the 1D chain is decaying with an exponent $\alpha=0.2$ in Eq.~(\ref{Equ:Jij}). The torroidal coupling $\phi=\phi_{34}=3\pi$ has to be rescaled to zero to be removed from the target graph. The Fourier coefficients directly give the duration of the pulses $\{t_j\}$ within a time cycle of the simulation.}
        \label{fig:FourierFit6Ion}
    \end{figure}
 \end{center}

 By comparing Eq.~(\ref{eq:betaij}) with Eq.~(\ref{Equ:Fourier}), we construct a multi-pulse sequence for implementing the Fourier filtering function $F(\phi)$, with the following features:
\begin{enumerate}
    \item The number of single qubit phase gates (by $\hext$) $l$ in each block is $l = 2i+1$, where the number of Fourier terms in Eq.~(\ref{Equ:Fourier}) is $i+1$. 
    
    \item The pulse sequence in each block is anti-symmetric about the central $\hat{H}_{\mathrm{ext}}$ pulse. That is $t_{j} = t_{l-j}$ and $\pm\hat{H}_{\mathrm{int}_j} = \mp\hat{H}_{\mathrm{int}_{l-j}}$ for $j = 1,..., l-1$. This leads to the cancellation of all even order correction terms to the average Hamiltonian in Eq.~(\ref{eq:Heff_multipleBlock2}).
    
    \item The time intervals $\{t_j\}$ are proportional to the coefficients in the Fourier filter function : $t_j = T|a_j|/2$ for $j = 1, \dots, l - 1$ and $t_l = T|a_0|$, with the constraint that $\sum_{j=0}^l |a_j|=1$. This constraint can be relaxed for an efficient search for the Fourier coefficients at the expense of reducing all the couplings in the target lattice by a global rescaling factor. Numerically, we find that  $\beta_{i,i+m}=0.7$ allows us to find efficient solutions for up to $N=100$.
    
    \item A negative coefficient ($a_j < 0$) in Eq.~(\ref{Equ:Fourier}) is implemented by an $\hext$ pulse followed by $-\hint$.
    
    \item We choose the phase gates to be of equal duration $\tau$, except the central phase gate in each block which has to have a duration of $\tau'= \tautot - (l-1)\tau$. $\tau$ can be read-off from $W = \tau / \tautot$.
    
\end{enumerate}
In Fig. \ref{fig:FourierFit6Ion}, we show the Fourier filter function fit for engineering a $2 \times 3$ square lattice when $\alpha = 0.2$ in Eq.~(\ref{Equ:Jij}). The nearest neighbor couplings $N_1$, except the toroidal linkage ($J_{34}$) have a phase of $\phitot = \pi$. The $N_3$ couplings accumulate a phase of $5\pi$. Hence we require $F(\pi)/F(5\pi)=\frac{1}{3^\alpha}=0.80$ such that $N_1$ couplings (except $J_{34}$) are equal to $N_3$. The toroidal linkage $J_{34}$ ($\phitot = 3\pi$) and $N_5$ couplings ($\phitot = 7\pi$) are scaled to zero. We have introduced a global rescaling factor to all the target couplings, by setting $F(5\pi)=0.7$. The global rescaling of all the target couplings ensures an efficient Fourier fit with minimum number of parameters for at least up to $N=100$ spins.
\begin{center}
\begin{table}
\setlength{\tabcolsep}{2.8pt}
\renewcommand{\arraystretch}{1.3}
\begin{tabular}{c|c|c|c|c|c|c|}
\cline{2-7}
   & $W$ & $a_0$ & $a_1$ & $a_2$ & $a_3$ & $a_4$  \\
\hline
    \multicolumn{1}{|c|}{6 ion $\alpha = 0.2$} &0.142 & 0.385 & 0.0436 & 0.114 & 0.457 & -  \\
\hline
    \multicolumn{1}{|c|}{9 ion $\alpha = 0.2$} &0.099 & 0.241 &  0.204 & -0.094 & 0.126 & 0.334 \\
\hline
\end{tabular}
\caption{Fourier fit parameters for a $2\times 3$ and a $3\times 3$ square lattice when $\alpha = 0.2$ (see Eq. \ref{Equ:Jij}). From these coefficients, we find the duration of $\pm\hint$ pulses as $t_i = t_{l-i} = T |a_i|/2$ for $i = 1,.., l-1$ and $t_l = T |a_0|$. Here, $T=T_{\mathrm{cyc}}/2$.}

\label{table:FourierFit}
\end{table}
\end{center}
\noindent In Table \ref{table:FourierFit}, the Fourier fit parameters for engineering a $2\times3$ (listed in Row 1) and a $3 \times 3$ square lattices (listed in Row 2) are given for $\alpha = 0.2$. 
\section{\label{sec:Results}RESULTS AND DISCUSSION}
\subsection{\label{subsec:NumericalSimulations}Numerical simulations}
\begin{center}
   \begin{figure*}
        \includegraphics[width=0.9\textwidth]{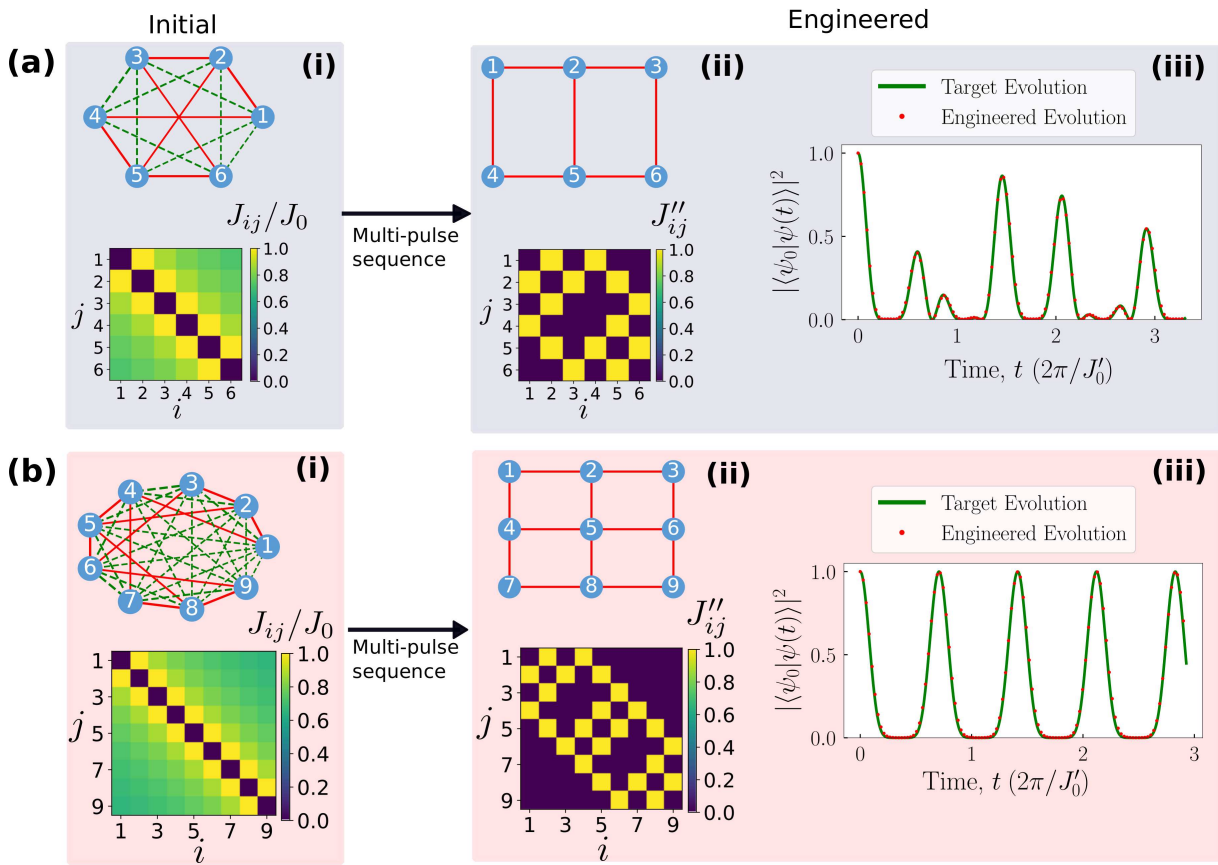}
        \caption{Benchmarking the dynamical Hamiltonian engineering against the target Hamiltonian for (a) a $2\times 3$ square lattice and (b) a $3\times 3$ square lattice. \textbf{(i)} The interaction graphs in the 1D ion chains with $\alpha = 0.2$ are shown, along with a 2D color plot of the couplings $J_{ij}$ vs $(i,j)$. The couplings are normalized to $J_0$ \textbf{(ii)} The engineered interaction graph closely resembles the target interaction graph of the square lattices with an RMS error (defined in the text) of $<0.1\%$. The engineered couplings $J''_{ij}=J'_{ij}/\mathrm{max}\{J'_{ij}\}$ are shown vs $(i,j)$ in the color plot. \textbf{(iii)} The evolution of the engineered lattice (red dots) is compared with the evolution of the target lattice (green curve). The system is initially prepared in (a) $|\psi_0\rangle = \mid\uparrow_1 \downarrow_2 \downarrow_3 \downarrow_4 \downarrow_5 \downarrow_6 \rangle$ and (b) $|\psi_0\rangle =\mid \uparrow_1 \downarrow_2  \downarrow_3 \downarrow_4 \downarrow_5 \downarrow_6\downarrow_7\downarrow_8\downarrow_9 \rangle$ state. The probability of measuring the state of the system in its initial state is shown over time. The evolution of the system under multi-pulse sequence replicates the evolution of the target lattice.}
        \label{fig:6ion_evolution_alpha0.2}
    \end{figure*}
 \end{center}

Applying the tools described above, we successfully reproduce the spin-spin interaction graph of the target square lattice at evolution times $t = nT_{\mathrm{cyc}}$, $n=1,2,\cdots$, starting with the initial long range couplings with an exponent $\alpha=0.2$. Figs.~\ref{fig:6ion_evolution_alpha0.2}a and \ref{fig:6ion_evolution_alpha0.2}b illustrate the results for a $2\times 3$ and a $3\times 3$ square lattices, respectively. The engineered interaction matrix formed by the couplings $\{J_{ij}\}$ matches the target interaction matrix of the 2D square lattice with an RMS error of $<0.1\%$. Here we define the RMS error as $\sqrt{\sum_{ij}(J^{\prime}_{ij} - J^{\prime}_{ij}(\mathrm{Target}))^2}/\sum_{ij}|J^{\prime}_{ij}(\mathrm{Target})|$, where $J^{\prime}_{ij}(\mathrm{Target})$ denote the couplings in the ideal lattice. In Fig.~\ref{fig:6ion_evolution_alpha0.2}, we also compare spin dynamics under the engineered lattice (red dots) with that of the ideal target (green curve), and find excellent agreement. Here the systems are initially prepared in $|\psi_0\rangle =\mid \uparrow_1 \downarrow_2 \downarrow_3 \downarrow_4 \downarrow_5 \downarrow_6 \rangle$ for $N=6$ and $|\psi_0\rangle =\mid \uparrow_1\downarrow_2 \downarrow_3 \downarrow_4 \downarrow_5 \downarrow_6\downarrow_7\downarrow_8\downarrow_9 \rangle$ for $N=9$ at $t=0$, when the pulse sequence is turned on. The probability of the system being in $|\psi_0\rangle$  (Figs.~\ref{fig:6ion_evolution_alpha0.2}a(iii) and ~\ref{fig:6ion_evolution_alpha0.2}b(iii)) follows the expected dynamics of the ideal 2D square lattice. These numerical simulations were performed using the time dependent master equation solver based on QUTIP python package \citep{Johansson2012,Johansson2013}.
 
The near-perfect matching of the target and engineered spin dynamics in Fig.~\ref{fig:6ion_evolution_alpha0.2} indicates small intrinsic errors. These errors arise from the Fourier fit in Eq.~(\ref{Equ:Fourier}) and numerical rounding error in the time interval values $\{t_i, \tau_i\}$. However, additional error may creep into an experimental realization due to imperfect single qubit gates. In our numerical simulation, we find that RMS error between the target and engineered interaction matrices scales linearly with single qubit phase error, with $1\%$ error in single qubit phase (in each pulse of $\hext$) contributing to approximately $1.2\%$ error in $J'_{ij}$.

\begin{center}
   \begin{figure}
        \includegraphics[width = 0.95\columnwidth]{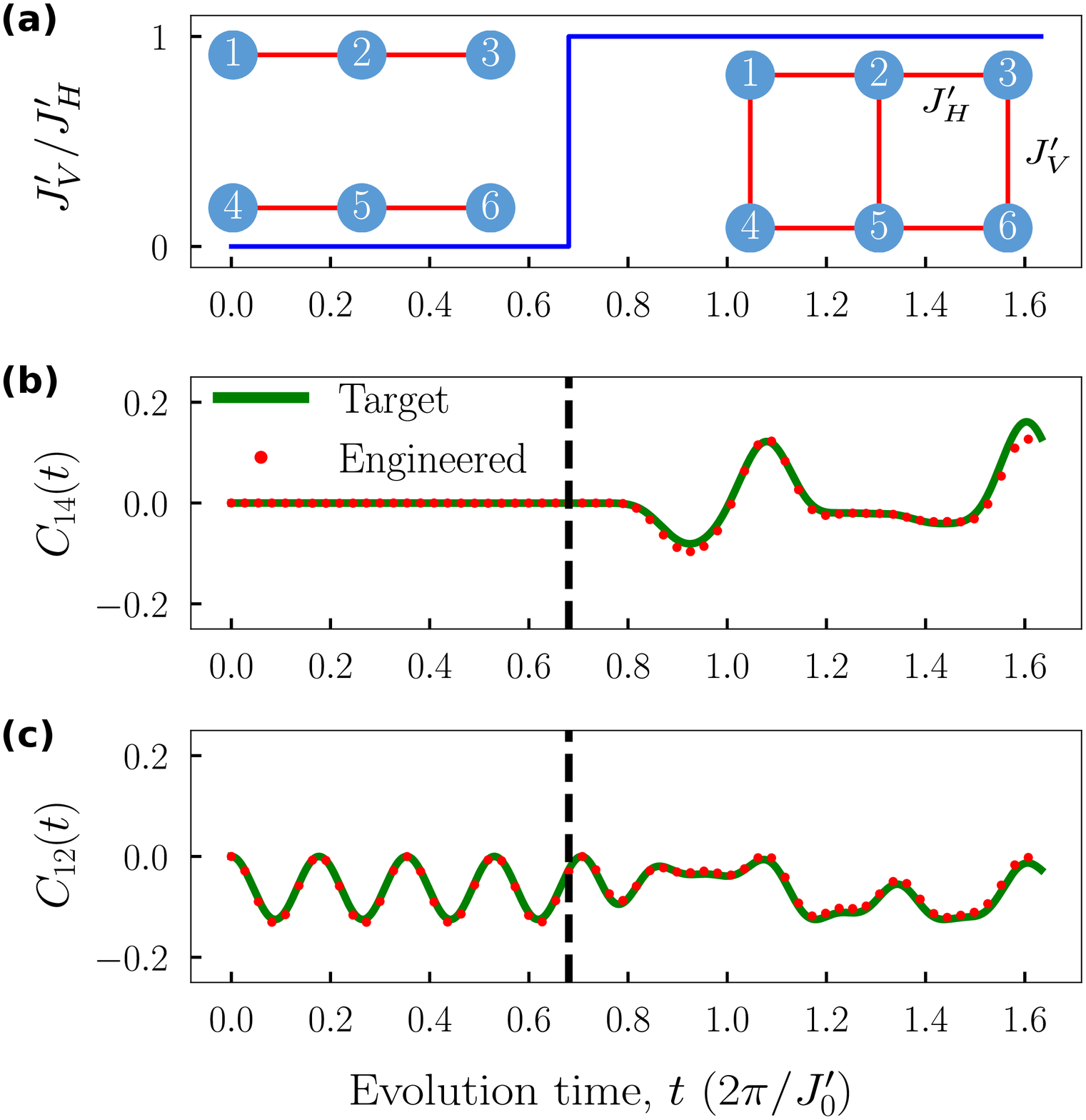}
        \caption{Dynamic changing of the target Hamiltonian. The Fourier filtering function, $F(\phi)$ in Eq.~(\ref{Equ:Fourier}) can be updated at each time cycle to realize a dynamically changing target Hamiltonian. For example, here we perform a quench from two decoupled chains of 3 spins each into a $2\times 3$ square lattice.  a) the correlation measure between Spin 1 and 4 defined by ${C_{14}(t) = \langle S_z^1S_z^4\rangle - \langle S_z^1\rangle\langle S_z^4\rangle}$ 
b) the correlation measure between Spin 1 and 2 defined by ${C_{12}(t) = \langle S_z^1S_z^2\rangle - \langle S_z^1\rangle\langle S_z^2\rangle}$. The spin-spin correlations between the previously uncoupled chains start to build up after the quench, indicated by the dashed line. The engineered dynamics follow closely to the ideal target dynamics.}\label{fig:Quench}
    \end{figure}
 \end{center}

A crucial feature of the protocol presented here is the ability to \textit{dynamically} change the Hamiltonian within the same symmetry class, by changing the Fourier coefficients in Eq.~(\ref{Equ:Fourier}). This enables simulation of many-body dynamics, such as quantum quench experiments that are hard to simulate numerically. As an example, we show a quench from two decoupled chains of 3 spins each into the $2\times 3$ square lattice plaquette in Fig. \ref{fig:Quench}. To simulate the decoupled chains, $N_3$ couplings are set to zero (see Fig.~\ref{fig:FourierFit6Ion}) in estimating the Fourier filtering function and hence the multi-pulse sequence. The spin-spin correlations between the previously uncoupled chains start to build up after the quench. We show the engineered dynamics (red dots) of the two point correlation functions $C_{12}(t)$ between spins 1 and 2 and $C_{14}(t)$ between spins 1 and 4. They follow closely to the ideal target dynamics (green line). 

\subsection{\label{subsec:ProposalExperimentalImplementation}Proposal for experimental implementation}

The experimental implementation of the multi-pulse scheme can be achieved in a trapped ion system such as $^{171}$Yb$^+$ trapped in a radio-frequency (Paul) trap. When the confining potential is sufficiently anisotropic, laser-cooled ions will form a linear chain \cite{Schiffer1993}. The hyperfine states $|0\rangle = |^2S_{1/2}, F = 0 ,m_F = 0 \rangle$ and $|1\rangle = |^2S_{1/2}, F = 1,m_F = 0\rangle$ form the qubit states for this ion \cite{Olmschenk2007}. 

The flip-flop Hamiltonian $\hat{H}_{\mathrm{int}}$ in Eq.~(\ref{Equ:H_XY}) can be simulated \cite{Jurcevic2014, Richerme2014} by global Raman laser beat-notes using the M{\o}lmer-S{\o}rensen scheme \cite{Molmer1999} for generating phonon-mediated spin-spin interactions. When the M{\o}lmer-S{\o}rensen detuning is tuned close to center of mass phonon mode, a long range interaction in the form of Eq.~(\ref{Equ:Jij}) can be obtained. The global sign of $\hat{H}_{\mathrm{int}}$ can be flipped by changing the M{\o}lmer-S{\o}rensen detuning, with additional detunings improving the  accuracy (see Appendix \ref{sec:Hint_to_minusHint} for details). 

The field gradient in $\hext$ can be implemented with laser beams imprinting AC Stark shifts, and by spatially modulating the laser intensity using a Spatial Light Modulator (SLM) or an Acousto Optic Deflector (AOD). Another potentially easier experimental implementation will be to combine a global tightly focused laser beam with additional relatively low power beams created by an AOD. The global beam propagating along the axis of the ion chain can be focused before hitting the ions, such that its intensity varies linearly on the ion chain. The jumps in the gradient field can be added by beams created by an SLM or AOD and shining on the ions from the transverse direction. A $100$ mW laser beam propagating along the ion chain, detuned from the $^{171}\mathrm{Yb}^+$ $^2S_{1/2}-^2P_{1/2}$ resonance by $10^5$ natural line-widths, and focused to approx. 2 microns will create a two-photon differential AC Stark shift gradient ($\omega_0$) of approximately 1 MHz. Thus, the total time ($\tautot$) that the Stark shifting beam is shining on the ions in a time cycle can be limited to a few microseconds, minimizing spontaneous emission errors.
\subsection{\label{subsec:Discussion}Discussions}

\begin{center}
   \begin{figure}
        \includegraphics[width = 0.9\columnwidth]{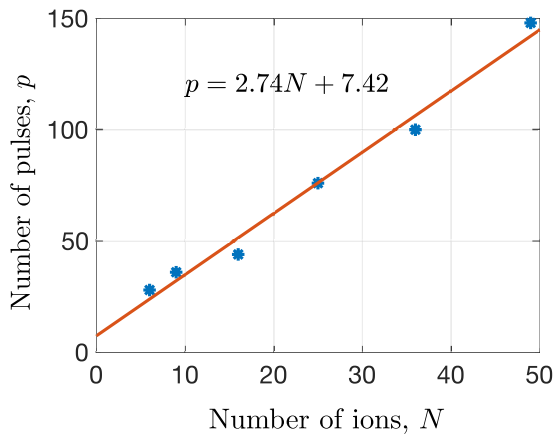}
        \caption{Number of pulses (includes $\hext$ and $\pm\hint$) in a time cycle, $T_\mathrm{cyc}$ required for engineering $m \times m$ square lattices as a function of the number of ions, $N = m^2$ in a 1D chain.}\label{fig:PulsevsIonNum}
    \end{figure}
 \end{center}

The Hamiltonian engineering protocol presented here makes efficient use of Fourier filtering to achieve a desired spin-spin interaction topology. For an arbitrary target lattice, $\mathcal{O}(N^2)$ Fourier coefficients, hence number of pulses, will be needed. However, in presence of common symmetries between the target lattice and the external field gradient, the number of pulses are reduced drastically. For example, the rectangular lattices presented here need $\mathcal{O}(N)$ pulses (Fig. \ref{fig:PulsevsIonNum}), as our chosen field gradient (Fig.~\ref{fig:labeling/gradient_scheme}b) creates the same phase tag for the class $N_d$, except a small number of ($\mathcal{O}(\sqrt{N})$) torroidal linkages. The number of Fourier coefficients are further reduced, approximately by a factor of 2, by using $\pm\hint$ instead of $\hint$ only. This is because all of the interactions for which $\phi_{ij}=2n\pi$ in a time cycle are canceled automatically (Eq.~(\ref{eq:Heff_twoBond3})) and hence are not required to be included in estimating the Fourier filtering function.  

The engineered interactions in the target 2D lattice will become weaker for a given $J_0$ in the original 1D chain, as the system size increases. This is because of the following reasons. The average Hamiltonian theory employed here works when $J_0 T_{\mathrm{cyc}}\ll 1$. Due to the linear scaling of the number of pulses in a cycle with $N$ (Fig.~\ref{fig:PulsevsIonNum}), $T_{\mathrm{cyc}}$ is expected to scale linearly, and hence the initial coupling $J_0$ has to be reduced linearly with increasing $N$. We may have to further scale some couplings down to match the target interaction graph, as the interactions are decaying with distance in the original 1D lattice. For example, the $N_1$ couplings (except the torroidal linkages) have to be scaled down by a factor of $1/m^\alpha$ compared to the $N_m$ couplings for an $m'\times m$ square lattice, as demonstrated in Fig.~\ref{fig:labeling/gradient_scheme} and Fig.~\ref{fig:FourierFit6Ion}. For the results presented here with $N=6$ and $N=9$ ions, we have chosen $\alpha=0.2$, which is experimentally realizable in current systems, and provides sufficiently strong target interactions. However, for $\alpha>0$, the couplings will scale down with increasing $N$, and hence longer simulation times are necessary as the system size increases.  We estimate a target coupling of $2\pi\times 300$ Hz in a $2\times 3$ square lattice. Since the separation between the vibrational normal modes in the ion chain will decrease with increasing $N$, the coupling strength $J_0$ will have to scale down accordingly in order to avoid direct excitation of phonons that limit the validity of a spin-only Hamiltonian.  
While trapped ion qubits have long single qubit coherence time \cite{Kim2010}, scaling the simulation to a large number of spins where classical computation of dynamics may be intractable will require isolating experimental noise sources, such as intensity fluctuations of the global M{\o}lmer-S{\o}rensen and Stark shifting laser beams, and drifts in the collective phonon mode frequencies. 
 
In a small chain of ions, the couplings are inhomogeneous. The errors due to the inhomogeneity can be mitigated by using an anharmonic trapping potential for the ions \cite{Lin2009} and spatially modulating the global M{\o}lmer-S{\o}rensen laser beams to increase the homogeneity of the couplings. The errors can also be reduced at the expense of increasing the number of pulses within a cycle and using a field gradient that breaks the symmetry between interactions belonging to a class $N_d$.

As the number of ions $N$ increases, the spacing between the vibrational modes decreases. This will make it harder to engineer $-\hint$ with a single global M{\o}lmer-S{\o}rensen beam that is detuned close to the center of mass mode. The accuracy of engineering $-\hint$ is enhanced by introducing additional global beams with M{\o}lmer-S{\o}rensen detuning near the neighboring phonon modes (see Appendix \ref{sec:Hint_to_minusHint}). Engineering $-\hint$ for very large $N$ will require either a large number of global M{\o}lmer-S{\o}rensen beams to cancel the effect of multiple modes, or reducing the overall intensity of laser beams resulting in a reduction in the strength of interactions ($J_0$) in the 1D chain. Engineering all the interactions via Fourier filtering alone (by using $\hint$ only instead of $\pm\hint$) may become experimentally preferable at the expense of a longer pulse sequence (and longer $T_\mathrm{cyc}$).

\section{\label{sec:conclusion}Conclusions and Outlook}
In this work, we proposed a hybrid analog-digital simulation protocol that leverages the long range spin-spin interactions in a linear chain of trapped ions to engineer a 2D lattice topology. This protocol efficiently engineers certain target interaction topology, such as 2D square lattices, by Fourier engineering analogous to holography. Our work opens up opportunities to identify classes of interaction graphs that can be simulated efficiently. 

The scheme needs only global beams creating spin-spin interactions and a single laser beam with intensity gradient. Unlike a full digital quantum simulator, this scheme does not require individual precise two qubit gates. An attractive feature of this protocol is the dynamical engineering of Hamiltonians, as we have demonstrated by a quench experiment. The dynamical engineering enables investigation of a range of many-body phenomena of active research interest, such as quantum quench and transport and dynamical phase transitions. The protocol is feasible with a moderately large system of tens of spins in existing trapped ion experiments. 
\section*{\label{sec:ACKNOWLEDGEMENTS}ACKNOWLEDGEMENTS}
We thank Industry Canada and University of Waterloo for financial assistance. FR and SM have been in part financially supported by Institute for Quantum Computing. This work is partially supported by a cooperative agreement with the Army Research Laboratory (W911NF-17-2-0117). AA would like to thank A. Pines and P. Cappellaro for insightful discussions. FR, SM, C-YS, NK, and RI acknowledge valuable discussions with Yi-Hong Teoh. 


\appendix

\section{SWITCHING BETWEEN $\hat{H}_{\mathrm{int}}$ AND $-\hat{H}_{\mathrm{int}}$}\label{sec:Hint_to_minusHint}

The phonon-mediated spin-spin interactions in an ion chain can be generated using off-resonant optical forces \cite{Molmer1999, Milburn2000, Leibfried2003, Porras2004}. For instance, two pairs of counter-propagating laser beams with a wave vector component $\Delta k$ perpendicular to the ion chain can induce transitions between the spin states and excite transverse vibrational phonon modes. When the optical beat-notes created by the lasers satisfy $\omega = \omega_\mathrm{s} \pm \mu$, with $\omega_\mathrm{m} \approx \mu \ll \omega_\mathrm{s}$, effective Ising interactions are created \citep{Molmer1999}. Here, $\omega_{\mathrm s}$ is the frequency difference between the spin states and $\omega_\mathrm{m}$ are the frequency of collective vibrational modes ($m=1,2,\cdots,N$). The Ising coupling strength $J_{ij}$ between spins $i$ and $j$ are given by,
\begin{equation}\label{eq:JijOptical}
    J_{ij} = \Omega_i \Omega_j \sum_{m} \frac{\eta_{i,m} \eta_{j,m} \omega_m}{{\mu}^2 - {\omega_m}^2}, 
\end{equation}
where $\Omega_i$ is the single ion Rabi frequency and $\eta_{i,m} = b_{i,m} \Delta k \sqrt{\hbar /2M\omega_m}$ is the Lamb-Dicke parameter indicating the coupling between the ion $i$ and phonon mode $m$. The Lamb-Dicke parameter depends on the mass of the ion $M$ and the eigenvector $\{b_{i,m}\}$ of the $m^\mathrm{th}$ normal mode. By applying an additional external field to the Ising Hamiltonian, the flip-flop or the XY Hamiltonian ($\hint$) is obtained \citep{Richerme2014,Jurcevic2014}.

To achieve long range interactions in $\hint$ falling-off as Eq.~(\ref{Equ:Jij}) and $\alpha\approx 0$, the beat-notes are chosen such that $\mu=\mu_1 = \omega_{\mathrm{COM}} + \delta_1$. Here, $\omega_\mathrm{COM}$ is the frequency of the center of mass (COM) mode, and the detuning $\delta_1>0$ is chosen such that the COM phonons are adiabatically eliminated. To switch between $\hat{H}_{\mathrm{int}}$ and $-\hat{H}_{\mathrm{int}}$, the sign of the detuning can be reversed, that is $\mu_1 = \omega_{\mathrm{COM}} + \delta_1 \rightarrow \mu^{\prime}_1 = \omega_{\mathrm{COM}} - \delta^{\prime}_1$ where $\delta^{\prime}_1>0$. When $\mu = \mu_1$, the COM mode mainly contributes to $J_{ij}$ with small contributions from other vibrational modes, mainly the tilt mode (the second vibrational mode):
\begin{equation}\label{eq:Jij_delta+}
J_{ij}[\delta_1] \propto \frac{(\eta_{\mathrm{COM}}\Omega_1)^2}{\delta_1} + \frac{(\eta_{\mathrm{tilt}}\Omega_1)^2}{\Delta + \delta_1},
\end{equation}
where $\Delta = \omega_{\mathrm{COM}} - \omega_{\mathrm{tilt}}$. When $\mu =\mu^{\prime}_1$,
\begin{equation}\label{eq:Jij_delta-}
J_{ij}[\delta^{\prime}_1]  \propto  -\frac{(\eta_{\mathrm{COM}}\Omega_1)^2}{\delta^{\prime}_1} + \frac{(\eta_{\mathrm{tilt}}\Omega_1)^2}{\Delta - \delta^{\prime}_1},
\end{equation}
which can be rearranged into
\begin{equation}\label{eq:Jij_delta-2}
J_{ij}[\delta^{\prime}_1] =  - J_{ij}[\delta_1] + \left(\frac{(\eta_{\mathrm{tilt}}\Omega_1)^2}{\Delta + \delta^{\prime}_1} + \frac{(\eta_{\mathrm{tilt}}\Omega_1)^2}{\Delta - \delta^{\prime}_1}\right),
\end{equation}
assuming $\delta_1 = \delta^{\prime}_1$. Eq. \ref{eq:Jij_delta-2} shows $J_{ij}[\delta_1] =  -J_{ij}[\delta^{\prime}_1]$ if the contribution from the second term $\left(\frac{(\eta_{\mathrm{tilt}}\Omega_1)^2}{\Delta + \delta^{\prime
}_1} + \frac{(\eta_{\mathrm{tilt}}\Omega_1)^2}{\Delta - \delta^{\prime}_1}\right)$ is canceled. This can be accomplished by applying a second pair of Raman beams with $\mu_2 = \omega_{\mathrm{tilt}} - \delta_2$, red-detuned from the tilt mode and chosen so that $\delta_2 < \delta^{\prime}_1$. This brings in another contribution from the tilt mode that scales as $-\left(\eta_{\mathrm{tilt}}\Omega_{2}\right)^2 / \delta_2$. One can optimize for the lasers parameters, i.e., Rabi frequencies and detunings, so that the coupling profiles for $\hint$ and $-\hint$ match as closely as possible. That is the difference between the corresponding coupling profiles defined by
$\Delta J_{ij} = |J_{ij}[\delta_1]-(-J_{ij}[\delta^{\prime}_1,\delta_2])|$ is negligible (note $J_{ij}[\delta^{\prime}_1,\delta_2])<0$).

\begin{center}
   \begin{figure}
        \includegraphics[width=0.9\columnwidth]{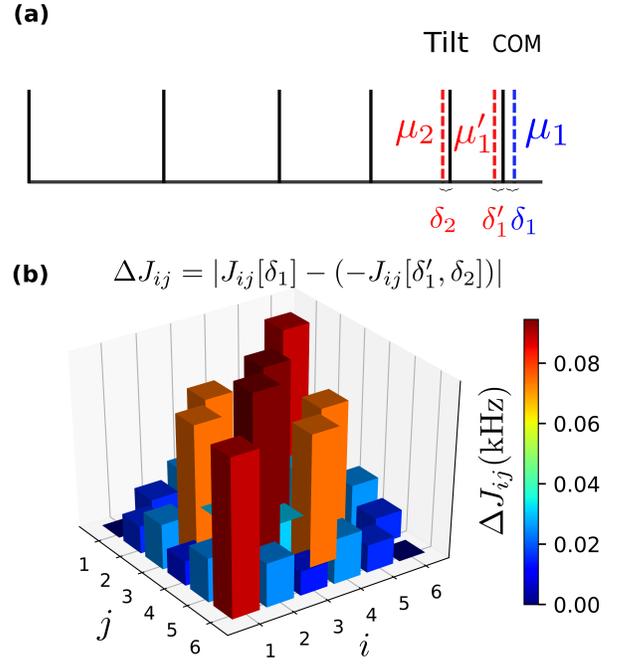}
        \caption{A set of M{\o}lmer-S{\o}rensen parameters to reduce errors when switching between $+\hat{H}_{\mathrm{int}}$ and $-\hat{H}_{\mathrm{int}}$ for a chain of 6 ions when $\alpha = 0.2$. a) Transverse normal mode spectrum (black solid lines) for $N = 6$ ions with the trap axial frequency 1.7 MHz and the transverse COM frequency 5 MHz. The lasers frequencies and detunings required to implement $+\hat{H}_{\mathrm{int}}$ and $-\hat{H}_{\mathrm{int}}$ are shown. b) The bar chart showing $\Delta J_{ij} = |J_{ij}[\delta_1] - (-J_{ij}[\delta^{\prime}_1, \delta_2])|$, the difference between $J_{ij}[\delta_1]$ and $J_{ij}[\delta^{\prime}_1, \delta_2]$ coupling profiles.
The bar chart indicates up to 1.9$\%$ error when switching from $\hat{H}_{\mathrm{int}}$ to $-\hat{H}_{\mathrm{int}}$ during the experiment.}
        \label{fig:Jij_6ion_experimental}
    \end{figure}
 \end{center}

In Fig. \ref{fig:Jij_6ion_experimental} we show a set of experimental parameters for a chain of 6 ions when $\alpha = 0.2$. To implement $+\hat{H}_{\mathrm{int}}$, a Raman beat-note with a global sideband Rabi frequency $\eta_{\mathrm{COM}}\Omega_1 = 2\pi \times 18$ kHz and $\delta_1= 2\pi \times 55$ kHz is blue-detuned from the COM mode (Fig. \ref{fig:Jij_6ion_experimental}a). This results in the coupling profile $J_{ij}[\delta_1]$ with the nearest neighbor coupling constant $J_0 \approx 2\pi \times 0.520$ kHz. To implement $-\hat{H}_{\mathrm{int}}$, two Raman beat-notes with $\delta^{\prime}_1 = 2\pi \times 45$ kHz and $\delta_2 = 2\pi \times 12.3$ kHz are red-detuned from the COM and tilt modes (Fig. \ref{fig:Jij_6ion_experimental}a). The sideband Rabi frequencies are taken to be $\eta_{\mathrm{COM}}\Omega^{\prime}_1 = 2\pi \times 15$ kHz and $\eta_{\mathrm{tilt}}\Omega_2 = 2\pi \times 4.1$ kHz, respectively, where $\eta_{\mathrm{COM}}\Omega^{\prime}_1$ is associated with $\mu^{\prime}_1$. These parameters result in the coupling profile $J_{ij}[\delta^{\prime}_1, \delta_2]$. The bar chart corresponding to $\Delta J_{ij} = J_{ij}[\delta_1] - (-J_{ij}[\delta^{\prime}_1, \delta_2])$ is shown in Fig. \ref{fig:Jij_6ion_experimental}b.
This bar chart indicates up to 1.9$\%$ error when switching between $\hat{H}_{\mathrm{int}}$ to $-\hat{H}_{\mathrm{int}}$ during the experiment.   
\section{LABELING/FIELD GRADIENT SCHEME FOR 9 IONS}\label{sec:9-ions}

\begin{figure}[h]
\begin{center}        \includegraphics[width = 0.9\columnwidth]{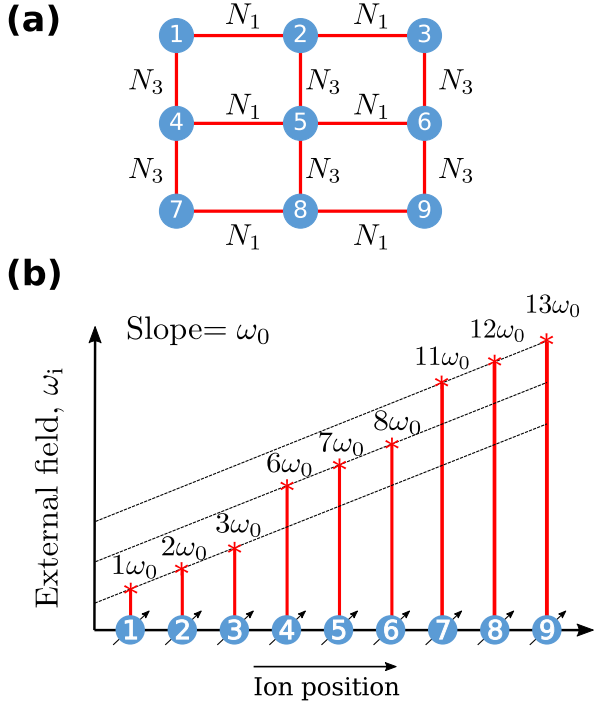}
        \caption{The labeling/field gradient scheme for engineering a $3\times 3$ square lattice. a) The ions in the 1D chain can be labeled in a way so that only $N_1$ and $N_3$ couplings are required for engineering the square lattice. The $N_1$ couplings form the horizontal bonds of the square lattice while $N_3$ couplings form the vertical bonds. b) The proposed field gradient scheme for engineering a $3\times3$ square lattice from a 1D chain of 9 ions. The external field profile corresponding to $\omega_i$ increases linearly along the chain of ions with two jumps of $+3\omega_0$ to assign distinct phase tags to $J_{34}$ and $J_{56}$ forming the toroidal linkage between the edges of the lattice.}
        \label{fig:9ion_labelling_gradient}
\end{center}
\end{figure}

Fig. \ref{fig:9ion_labelling_gradient}a illustrates the labeling scheme for engineering a $3\times 3$ square lattice from a 1D chain of 9 ions. The class A interactions in the ion chain network consists of $N_1$ and $N_3$ interactions. The $\{J_{34} ,J_{56} \in N_1\}$ should be excluded as they form toroidal linkages at the horizontal edges of the square lattice and should be set to zero to engineer the target geometry. Fig. \ref{fig:9ion_labelling_gradient}b illustrates the external field gradient scheme proposed for engineering a $3\times 3$ square lattice. The external field profile increases linearly across the ion chain with two jumps of $+3\omega_0$ to assign a distinct tag to $J_{34}$ and $J_{56}$ interactions. Similar to the 6-ion chain, one can adjust $\omega_0 \tau_{\mathrm{tot}} = \pi$ so that $\phi^{A}_{{ij}_{\mathrm{tot}}} = (2n-1)\pi$ and $\phi^{B}_{{ij}_{\mathrm{tot}}} =2n\pi$ except for $N_5, N_7 \in$ class B. 
\section{FOURIER FILTER FUNCTION FIT PARAMETERS}\label{sec:fourier_fit_paras}
\begin{table}[H]
\begin{center}
\setlength{\tabcolsep}{3pt}
\renewcommand{\arraystretch}{1.3}
\begin{tabular}{|c|c|c|c|c|}
\cline{2-5}
\multicolumn{1}{c|}{}  & 16 ions & 25 ions & 36 ions & 49 ions  \\
\hline
$W$       &  0.0833337 &  0.0499969 &  0.0384582 &  0.026316 \\
\hline
$a_{0}$  &   0.204912 &   0.120175 &  0.0912132 &  0.062292 \\
\hline
$a_{1}$   &   0.230988 &   0.162846 &   0.145688 &  0.104796 \\
\hline
$a_{2}$   & -0.0486727 &  0.0136359 &  0.0598559 &  0.055799 \\
\hline
$a_{3}$   &  -0.039712 &  -0.047381 & -0.0179331 &  0.003215 \\
\hline
$a_{4}$  &   0.204859 &  0.0381538 & -0.0376196 & -0.025686 \\
\hline
$a_{5}$   &   0.270857 &   0.169724 &   0.011602 & -0.016130 \\
\hline
$a_{6}$  &          - &   0.191738 &  0.0940426 &  0.026047 \\
\hline
$a_{7}$   &          - &  0.0673324 &   0.150076 &  0.077484 \\
\hline
$a_{8}$   &          - & -0.0810888 &   0.137711 &  0.109961 \\
\hline
$a_{9}$  &          - &  -0.107924 &  0.0605449 &  0.105051 \\
\hline
$a_{10}$  &          - &          - &  -0.034126 &  0.063949 \\
\hline
$a_{11}$  &          - &          - & -0.0884624 &  0.006843 \\
\hline
$a_{12}$ &          - &          - & -0.0711247 & -0.037818 \\
\hline
$a_{13}$ &          - &          - &          - & -0.048081 \\
\hline
$a_{14}$  &          - &          - &          - & -0.020459 \\
\hline
$a_{15}$  &          - &          - &          - &  0.028583 \\
\hline
$a_{16}$ &          - &          - &          - &  0.070986 \\
\hline
$a_{17}$ &          - &          - &          - &  0.082330 \\
\hline
$a_{18}$  &          - &          - &          - &  0.054488 \\
\hline
\end{tabular}
\caption{Fourier filter function fit parameters $\{W, a_i\}$ for $m\times m$ square lattices when $m = 3,4,5,6,7$.}
\label{table:FourierFit}
\end{center}
\end{table} 
\begin{center}
	\begin{tikzpicture}
        \fill (-3,-.25) -- ++(3,.2mm) -- ++(3,-.2mm) -- ++(-3,-.1mm) -- cycle;
    \end{tikzpicture}
\end{center}
\bibliography{references}

\begin{thebibliography}{65}%
\makeatletter
\providecommand \@ifxundefined [1]{%
 \@ifx{#1\undefined}
}%
\providecommand \@ifnum [1]{%
 \ifnum #1\expandafter \@firstoftwo
 \else \expandafter \@secondoftwo
 \fi
}%
\providecommand \@ifx [1]{%
 \ifx #1\expandafter \@firstoftwo
 \else \expandafter \@secondoftwo
 \fi
}%
\providecommand \natexlab [1]{#1}%
\providecommand \enquote  [1]{``#1''}%
\providecommand \bibnamefont  [1]{#1}%
\providecommand \bibfnamefont [1]{#1}%
\providecommand \citenamefont [1]{#1}%
\providecommand \href@noop [0]{\@secondoftwo}%
\providecommand \href [0]{\begingroup \@sanitize@url \@href}%
\providecommand \@href[1]{\@@startlink{#1}\@@href}%
\providecommand \@@href[1]{\endgroup#1\@@endlink}%
\providecommand \@sanitize@url [0]{\catcode `\\12\catcode `\$12\catcode
  `\&12\catcode `\#12\catcode `\^12\catcode `\_12\catcode `\%12\relax}%
\providecommand \@@startlink[1]{}%
\providecommand \@@endlink[0]{}%
\providecommand \url  [0]{\begingroup\@sanitize@url \@url }%
\providecommand \@url [1]{\endgroup\@href {#1}{\urlprefix }}%
\providecommand \urlprefix  [0]{URL }%
\providecommand \Eprint [0]{\href }%
\providecommand \doibase [0]{http://dx.doi.org/}%
\providecommand \selectlanguage [0]{\@gobble}%
\providecommand \bibinfo  [0]{\@secondoftwo}%
\providecommand \bibfield  [0]{\@secondoftwo}%
\providecommand \translation [1]{[#1]}%
\providecommand \BibitemOpen [0]{}%
\providecommand \bibitemStop [0]{}%
\providecommand \bibitemNoStop [0]{.\EOS\space}%
\providecommand \EOS [0]{\spacefactor3000\relax}%
\providecommand \BibitemShut  [1]{\csname bibitem#1\endcsname}%
\let\auto@bib@innerbib\@empty
\bibitem [{\citenamefont {Aspuru-Guzik}\ and\ \citenamefont
  {Walther}(2012)}]{Aspuru2012}%
  \BibitemOpen
  \bibfield  {author} {\bibinfo {author} {\bibfnamefont {A.}~\bibnamefont
  {Aspuru-Guzik}}\ and\ \bibinfo {author} {\bibfnamefont {P.}~\bibnamefont
  {Walther}},\ }\href@noop {} {\bibfield  {journal} {\bibinfo  {journal}
  {Nature Physics}\ }\textbf {\bibinfo {volume} {8}},\ \bibinfo {pages} {285}
  (\bibinfo {year} {2012})}\BibitemShut {NoStop}%
\bibitem [{\citenamefont {Bloch}\ \emph {et~al.}(2012)\citenamefont {Bloch},
  \citenamefont {Dalibard},\ and\ \citenamefont {Nascimbene}}]{Bloch2012}%
  \BibitemOpen
  \bibfield  {author} {\bibinfo {author} {\bibfnamefont {I.}~\bibnamefont
  {Bloch}}, \bibinfo {author} {\bibfnamefont {J.}~\bibnamefont {Dalibard}}, \
  and\ \bibinfo {author} {\bibfnamefont {S.}~\bibnamefont {Nascimbene}},\
  }\href@noop {} {\bibfield  {journal} {\bibinfo  {journal} {Nature Physics}\
  }\textbf {\bibinfo {volume} {8}},\ \bibinfo {pages} {267} (\bibinfo {year}
  {2012})}\BibitemShut {NoStop}%
\bibitem [{\citenamefont {Blatt}\ and\ \citenamefont {Roos}(2012)}]{Blatt2012}%
  \BibitemOpen
  \bibfield  {author} {\bibinfo {author} {\bibfnamefont {R.}~\bibnamefont
  {Blatt}}\ and\ \bibinfo {author} {\bibfnamefont {C.~F.}\ \bibnamefont
  {Roos}},\ }\href@noop {} {\bibfield  {journal} {\bibinfo  {journal} {Nature
  Physics}\ }\textbf {\bibinfo {volume} {8}},\ \bibinfo {pages} {277} (\bibinfo
  {year} {2012})}\BibitemShut {NoStop}%
\bibitem [{\citenamefont {Cirac}\ and\ \citenamefont
  {Zoller}(2012)}]{Cirac2012}%
  \BibitemOpen
  \bibfield  {author} {\bibinfo {author} {\bibfnamefont {J.~I.}\ \bibnamefont
  {Cirac}}\ and\ \bibinfo {author} {\bibfnamefont {P.}~\bibnamefont {Zoller}},\
  }\href@noop {} {\bibfield  {journal} {\bibinfo  {journal} {Nature Physics}\
  }\textbf {\bibinfo {volume} {8}},\ \bibinfo {pages} {264} (\bibinfo {year}
  {2012})}\BibitemShut {NoStop}%
\bibitem [{\citenamefont {Bloch}(2005)}]{Bloch2005}%
  \BibitemOpen
  \bibfield  {author} {\bibinfo {author} {\bibfnamefont {I.}~\bibnamefont
  {Bloch}},\ }\href@noop {} {\bibfield  {journal} {\bibinfo  {journal} {Nature
  Physics}\ }\textbf {\bibinfo {volume} {1}},\ \bibinfo {pages} {23} (\bibinfo
  {year} {2005})}\BibitemShut {NoStop}%
\bibitem [{\citenamefont {Gross}\ and\ \citenamefont
  {Bloch}(2017)}]{Gross2017}%
  \BibitemOpen
  \bibfield  {author} {\bibinfo {author} {\bibfnamefont {C.}~\bibnamefont
  {Gross}}\ and\ \bibinfo {author} {\bibfnamefont {I.}~\bibnamefont {Bloch}},\
  }\href@noop {} {\bibfield  {journal} {\bibinfo  {journal} {Science}\ }\textbf
  {\bibinfo {volume} {357}},\ \bibinfo {pages} {995} (\bibinfo {year}
  {2017})}\BibitemShut {NoStop}%
\bibitem [{\citenamefont {Buluta}\ and\ \citenamefont
  {Nori}(2009)}]{Buluta2009}%
  \BibitemOpen
  \bibfield  {author} {\bibinfo {author} {\bibfnamefont {I.}~\bibnamefont
  {Buluta}}\ and\ \bibinfo {author} {\bibfnamefont {F.}~\bibnamefont {Nori}},\
  }\href@noop {} {\bibfield  {journal} {\bibinfo  {journal} {Science}\ }\textbf
  {\bibinfo {volume} {326}},\ \bibinfo {pages} {108} (\bibinfo {year}
  {2009})}\BibitemShut {NoStop}%
\bibitem [{\citenamefont {Houck}\ \emph {et~al.}(2012)\citenamefont {Houck},
  \citenamefont {T{\"u}reci},\ and\ \citenamefont {Koch}}]{Houck2012}%
  \BibitemOpen
  \bibfield  {author} {\bibinfo {author} {\bibfnamefont {A.~A.}\ \bibnamefont
  {Houck}}, \bibinfo {author} {\bibfnamefont {H.~E.}\ \bibnamefont
  {T{\"u}reci}}, \ and\ \bibinfo {author} {\bibfnamefont {J.}~\bibnamefont
  {Koch}},\ }\href@noop {} {\bibfield  {journal} {\bibinfo  {journal} {Nature
  Physics}\ }\textbf {\bibinfo {volume} {8}},\ \bibinfo {pages} {292} (\bibinfo
  {year} {2012})}\BibitemShut {NoStop}%
\bibitem [{\citenamefont {Devoret}\ and\ \citenamefont
  {Schoelkopf}(2013)}]{Devoret2013}%
  \BibitemOpen
  \bibfield  {author} {\bibinfo {author} {\bibfnamefont {M.~H.}\ \bibnamefont
  {Devoret}}\ and\ \bibinfo {author} {\bibfnamefont {R.~J.}\ \bibnamefont
  {Schoelkopf}},\ }\href@noop {} {\bibfield  {journal} {\bibinfo  {journal}
  {Science}\ }\textbf {\bibinfo {volume} {339}},\ \bibinfo {pages} {1169}
  (\bibinfo {year} {2013})}\BibitemShut {NoStop}%
\bibitem [{\citenamefont {Korenblit}\ \emph {et~al.}(2012)\citenamefont
  {Korenblit}, \citenamefont {Kafri}, \citenamefont {Campbell}, \citenamefont
  {Islam}, \citenamefont {Edwards}, \citenamefont {Gong}, \citenamefont {Lin},
  \citenamefont {Duan}, \citenamefont {Kim}, \citenamefont {Kim} \emph
  {et~al.}}]{Korenblit2012}%
  \BibitemOpen
  \bibfield  {author} {\bibinfo {author} {\bibfnamefont {S.}~\bibnamefont
  {Korenblit}}, \bibinfo {author} {\bibfnamefont {D.}~\bibnamefont {Kafri}},
  \bibinfo {author} {\bibfnamefont {W.~C.}\ \bibnamefont {Campbell}}, \bibinfo
  {author} {\bibfnamefont {R.}~\bibnamefont {Islam}}, \bibinfo {author}
  {\bibfnamefont {E.~E.}\ \bibnamefont {Edwards}}, \bibinfo {author}
  {\bibfnamefont {Z.-X.}\ \bibnamefont {Gong}}, \bibinfo {author}
  {\bibfnamefont {G.-D.}\ \bibnamefont {Lin}}, \bibinfo {author} {\bibfnamefont
  {L.}~\bibnamefont {Duan}}, \bibinfo {author} {\bibfnamefont {J.}~\bibnamefont
  {Kim}}, \bibinfo {author} {\bibfnamefont {K.}~\bibnamefont {Kim}},  \emph
  {et~al.},\ }\href@noop {} {\bibfield  {journal} {\bibinfo  {journal} {New
  Journal of Physics}\ }\textbf {\bibinfo {volume} {14}},\ \bibinfo {pages}
  {095024} (\bibinfo {year} {2012})}\BibitemShut {NoStop}%
\bibitem [{\citenamefont {M{\o}lmer}\ and\ \citenamefont
  {S{\o}rensen}(1999)}]{Molmer1999}%
  \BibitemOpen
  \bibfield  {author} {\bibinfo {author} {\bibfnamefont {K.}~\bibnamefont
  {M{\o}lmer}}\ and\ \bibinfo {author} {\bibfnamefont {A.}~\bibnamefont
  {S{\o}rensen}},\ }\href@noop {} {\bibfield  {journal} {\bibinfo  {journal}
  {Physical Review Letters}\ }\textbf {\bibinfo {volume} {82}},\ \bibinfo
  {pages} {1835} (\bibinfo {year} {1999})}\BibitemShut {NoStop}%
\bibitem [{\citenamefont {Deng}\ \emph {et~al.}(2005)\citenamefont {Deng},
  \citenamefont {Porras},\ and\ \citenamefont {Cirac}}]{Deng2005}%
  \BibitemOpen
  \bibfield  {author} {\bibinfo {author} {\bibfnamefont {X.-L.}\ \bibnamefont
  {Deng}}, \bibinfo {author} {\bibfnamefont {D.}~\bibnamefont {Porras}}, \ and\
  \bibinfo {author} {\bibfnamefont {J.~I.}\ \bibnamefont {Cirac}},\ }\href@noop
  {} {\bibfield  {journal} {\bibinfo  {journal} {Physical Review A}\ }\textbf
  {\bibinfo {volume} {72}},\ \bibinfo {pages} {063407} (\bibinfo {year}
  {2005})}\BibitemShut {NoStop}%
\bibitem [{\citenamefont {Kim}\ \emph {et~al.}(2009)\citenamefont {Kim},
  \citenamefont {Chang}, \citenamefont {Islam}, \citenamefont {Korenblit},
  \citenamefont {Duan},\ and\ \citenamefont {Monroe}}]{Kim2009}%
  \BibitemOpen
  \bibfield  {author} {\bibinfo {author} {\bibfnamefont {K.}~\bibnamefont
  {Kim}}, \bibinfo {author} {\bibfnamefont {M.-S.}\ \bibnamefont {Chang}},
  \bibinfo {author} {\bibfnamefont {R.}~\bibnamefont {Islam}}, \bibinfo
  {author} {\bibfnamefont {S.}~\bibnamefont {Korenblit}}, \bibinfo {author}
  {\bibfnamefont {L.-M.}\ \bibnamefont {Duan}}, \ and\ \bibinfo {author}
  {\bibfnamefont {C.}~\bibnamefont {Monroe}},\ }\href@noop {} {\bibfield
  {journal} {\bibinfo  {journal} {Physical review letters}\ }\textbf {\bibinfo
  {volume} {103}},\ \bibinfo {pages} {120502} (\bibinfo {year}
  {2009})}\BibitemShut {NoStop}%
\bibitem [{\citenamefont {Kim}\ \emph {et~al.}(2010)\citenamefont {Kim},
  \citenamefont {Chang}, \citenamefont {Korenblit}, \citenamefont {Islam},
  \citenamefont {Edwards}, \citenamefont {Freericks}, \citenamefont {Lin},
  \citenamefont {Duan},\ and\ \citenamefont {Monroe}}]{Kim2010}%
  \BibitemOpen
  \bibfield  {author} {\bibinfo {author} {\bibfnamefont {K.}~\bibnamefont
  {Kim}}, \bibinfo {author} {\bibfnamefont {M.-S.}\ \bibnamefont {Chang}},
  \bibinfo {author} {\bibfnamefont {S.}~\bibnamefont {Korenblit}}, \bibinfo
  {author} {\bibfnamefont {R.}~\bibnamefont {Islam}}, \bibinfo {author}
  {\bibfnamefont {E.~E.}\ \bibnamefont {Edwards}}, \bibinfo {author}
  {\bibfnamefont {J.~K.}\ \bibnamefont {Freericks}}, \bibinfo {author}
  {\bibfnamefont {G.-D.}\ \bibnamefont {Lin}}, \bibinfo {author} {\bibfnamefont
  {L.-M.}\ \bibnamefont {Duan}}, \ and\ \bibinfo {author} {\bibfnamefont
  {C.}~\bibnamefont {Monroe}},\ }\href@noop {} {\bibfield  {journal} {\bibinfo
  {journal} {Nature}\ }\textbf {\bibinfo {volume} {465}},\ \bibinfo {pages}
  {590} (\bibinfo {year} {2010})}\BibitemShut {NoStop}%
\bibitem [{\citenamefont {Britton}\ \emph {et~al.}(2012)\citenamefont
  {Britton}, \citenamefont {Sawyer}, \citenamefont {Keith}, \citenamefont
  {Wang}, \citenamefont {Freericks}, \citenamefont {Uys}, \citenamefont
  {Biercuk},\ and\ \citenamefont {Bollinger}}]{Britton2012}%
  \BibitemOpen
  \bibfield  {author} {\bibinfo {author} {\bibfnamefont {J.~W.}\ \bibnamefont
  {Britton}}, \bibinfo {author} {\bibfnamefont {B.~C.}\ \bibnamefont {Sawyer}},
  \bibinfo {author} {\bibfnamefont {A.~C.}\ \bibnamefont {Keith}}, \bibinfo
  {author} {\bibfnamefont {C.-C.~J.}\ \bibnamefont {Wang}}, \bibinfo {author}
  {\bibfnamefont {J.~K.}\ \bibnamefont {Freericks}}, \bibinfo {author}
  {\bibfnamefont {H.}~\bibnamefont {Uys}}, \bibinfo {author} {\bibfnamefont
  {M.~J.}\ \bibnamefont {Biercuk}}, \ and\ \bibinfo {author} {\bibfnamefont
  {J.~J.}\ \bibnamefont {Bollinger}},\ }\href@noop {} {\bibfield  {journal}
  {\bibinfo  {journal} {Nature}\ }\textbf {\bibinfo {volume} {484}},\ \bibinfo
  {pages} {489} (\bibinfo {year} {2012})}\BibitemShut {NoStop}%
\bibitem [{\citenamefont {Islam}\ \emph {et~al.}(2013)\citenamefont {Islam},
  \citenamefont {Senko}, \citenamefont {Campbell}, \citenamefont {Korenblit},
  \citenamefont {Smith}, \citenamefont {Lee}, \citenamefont {Edwards},
  \citenamefont {Wang}, \citenamefont {Freericks},\ and\ \citenamefont
  {Monroe}}]{Islam2013}%
  \BibitemOpen
  \bibfield  {author} {\bibinfo {author} {\bibfnamefont {R.}~\bibnamefont
  {Islam}}, \bibinfo {author} {\bibfnamefont {C.}~\bibnamefont {Senko}},
  \bibinfo {author} {\bibfnamefont {W.}~\bibnamefont {Campbell}}, \bibinfo
  {author} {\bibfnamefont {S.}~\bibnamefont {Korenblit}}, \bibinfo {author}
  {\bibfnamefont {J.}~\bibnamefont {Smith}}, \bibinfo {author} {\bibfnamefont
  {A.}~\bibnamefont {Lee}}, \bibinfo {author} {\bibfnamefont {E.}~\bibnamefont
  {Edwards}}, \bibinfo {author} {\bibfnamefont {C.-C.}\ \bibnamefont {Wang}},
  \bibinfo {author} {\bibfnamefont {J.}~\bibnamefont {Freericks}}, \ and\
  \bibinfo {author} {\bibfnamefont {C.}~\bibnamefont {Monroe}},\ }\href@noop {}
  {\bibfield  {journal} {\bibinfo  {journal} {Science}\ }\textbf {\bibinfo
  {volume} {340}},\ \bibinfo {pages} {583} (\bibinfo {year}
  {2013})}\BibitemShut {NoStop}%
\bibitem [{\citenamefont {Richerme}\ \emph {et~al.}(2014)\citenamefont
  {Richerme}, \citenamefont {Gong}, \citenamefont {Lee}, \citenamefont {Senko},
  \citenamefont {Smith}, \citenamefont {Foss-Feig}, \citenamefont {Michalakis},
  \citenamefont {Gorshkov},\ and\ \citenamefont {Monroe}}]{Richerme2014}%
  \BibitemOpen
  \bibfield  {author} {\bibinfo {author} {\bibfnamefont {P.}~\bibnamefont
  {Richerme}}, \bibinfo {author} {\bibfnamefont {Z.-X.}\ \bibnamefont {Gong}},
  \bibinfo {author} {\bibfnamefont {A.}~\bibnamefont {Lee}}, \bibinfo {author}
  {\bibfnamefont {C.}~\bibnamefont {Senko}}, \bibinfo {author} {\bibfnamefont
  {J.}~\bibnamefont {Smith}}, \bibinfo {author} {\bibfnamefont
  {M.}~\bibnamefont {Foss-Feig}}, \bibinfo {author} {\bibfnamefont
  {S.}~\bibnamefont {Michalakis}}, \bibinfo {author} {\bibfnamefont {A.~V.}\
  \bibnamefont {Gorshkov}}, \ and\ \bibinfo {author} {\bibfnamefont
  {C.}~\bibnamefont {Monroe}},\ }\href@noop {} {\bibfield  {journal} {\bibinfo
  {journal} {Nature}\ }\textbf {\bibinfo {volume} {511}},\ \bibinfo {pages}
  {198} (\bibinfo {year} {2014})}\BibitemShut {NoStop}%
\bibitem [{\citenamefont {Jurcevic}\ \emph {et~al.}(2014)\citenamefont
  {Jurcevic}, \citenamefont {Lanyon}, \citenamefont {Hauke}, \citenamefont
  {Hempel}, \citenamefont {Zoller}, \citenamefont {Blatt},\ and\ \citenamefont
  {Roos}}]{Jurcevic2014}%
  \BibitemOpen
  \bibfield  {author} {\bibinfo {author} {\bibfnamefont {P.}~\bibnamefont
  {Jurcevic}}, \bibinfo {author} {\bibfnamefont {B.~P.}\ \bibnamefont
  {Lanyon}}, \bibinfo {author} {\bibfnamefont {P.}~\bibnamefont {Hauke}},
  \bibinfo {author} {\bibfnamefont {C.}~\bibnamefont {Hempel}}, \bibinfo
  {author} {\bibfnamefont {P.}~\bibnamefont {Zoller}}, \bibinfo {author}
  {\bibfnamefont {R.}~\bibnamefont {Blatt}}, \ and\ \bibinfo {author}
  {\bibfnamefont {C.~F.}\ \bibnamefont {Roos}},\ }\href@noop {} {\bibfield
  {journal} {\bibinfo  {journal} {Nature}\ }\textbf {\bibinfo {volume} {511}},\
  \bibinfo {pages} {202} (\bibinfo {year} {2014})}\BibitemShut {NoStop}%
\bibitem [{\citenamefont {Bohnet}\ \emph {et~al.}(2016)\citenamefont {Bohnet},
  \citenamefont {Sawyer}, \citenamefont {Britton}, \citenamefont {Wall},
  \citenamefont {Rey}, \citenamefont {Foss-Feig},\ and\ \citenamefont
  {Bollinger}}]{Bohnet2016}%
  \BibitemOpen
  \bibfield  {author} {\bibinfo {author} {\bibfnamefont {J.~G.}\ \bibnamefont
  {Bohnet}}, \bibinfo {author} {\bibfnamefont {B.~C.}\ \bibnamefont {Sawyer}},
  \bibinfo {author} {\bibfnamefont {J.~W.}\ \bibnamefont {Britton}}, \bibinfo
  {author} {\bibfnamefont {M.~L.}\ \bibnamefont {Wall}}, \bibinfo {author}
  {\bibfnamefont {A.~M.}\ \bibnamefont {Rey}}, \bibinfo {author} {\bibfnamefont
  {M.}~\bibnamefont {Foss-Feig}}, \ and\ \bibinfo {author} {\bibfnamefont
  {J.~J.}\ \bibnamefont {Bollinger}},\ }\href@noop {} {\bibfield  {journal}
  {\bibinfo  {journal} {Science}\ }\textbf {\bibinfo {volume} {352}},\ \bibinfo
  {pages} {1297} (\bibinfo {year} {2016})}\BibitemShut {NoStop}%
\bibitem [{\citenamefont {Qi}\ and\ \citenamefont {Zhang}(2011)}]{Qi2011}%
  \BibitemOpen
  \bibfield  {author} {\bibinfo {author} {\bibfnamefont {X.-L.}\ \bibnamefont
  {Qi}}\ and\ \bibinfo {author} {\bibfnamefont {S.-C.}\ \bibnamefont {Zhang}},\
  }\href@noop {} {\bibfield  {journal} {\bibinfo  {journal} {Reviews of Modern
  Physics}\ }\textbf {\bibinfo {volume} {83}},\ \bibinfo {pages} {1057}
  (\bibinfo {year} {2011})}\BibitemShut {NoStop}%
\bibitem [{\citenamefont {Haldane}\ and\ \citenamefont
  {Raghu}(2008)}]{Haldane2008}%
  \BibitemOpen
  \bibfield  {author} {\bibinfo {author} {\bibfnamefont {F.}~\bibnamefont
  {Haldane}}\ and\ \bibinfo {author} {\bibfnamefont {S.}~\bibnamefont
  {Raghu}},\ }\href@noop {} {\bibfield  {journal} {\bibinfo  {journal}
  {Physical review letters}\ }\textbf {\bibinfo {volume} {100}},\ \bibinfo
  {pages} {013904} (\bibinfo {year} {2008})}\BibitemShut {NoStop}%
\bibitem [{\citenamefont {Kitaev}(2003)}]{Kitaev2003}%
  \BibitemOpen
  \bibfield  {author} {\bibinfo {author} {\bibfnamefont {A.~Y.}\ \bibnamefont
  {Kitaev}},\ }\href@noop {} {\bibfield  {journal} {\bibinfo  {journal} {Annals
  of Physics}\ }\textbf {\bibinfo {volume} {303}},\ \bibinfo {pages} {2}
  (\bibinfo {year} {2003})}\BibitemShut {NoStop}%
\bibitem [{\citenamefont {Schmied}\ \emph {et~al.}(2011)\citenamefont
  {Schmied}, \citenamefont {Wesenberg},\ and\ \citenamefont
  {Leibfried}}]{Schmied2011}%
  \BibitemOpen
  \bibfield  {author} {\bibinfo {author} {\bibfnamefont {R.}~\bibnamefont
  {Schmied}}, \bibinfo {author} {\bibfnamefont {J.~H.}\ \bibnamefont
  {Wesenberg}}, \ and\ \bibinfo {author} {\bibfnamefont {D.}~\bibnamefont
  {Leibfried}},\ }\href@noop {} {\bibfield  {journal} {\bibinfo  {journal} {New
  Journal of Physics}\ }\textbf {\bibinfo {volume} {13}},\ \bibinfo {pages}
  {115011} (\bibinfo {year} {2011})}\BibitemShut {NoStop}%
\bibitem [{\citenamefont {Linke}\ \emph {et~al.}(2017)\citenamefont {Linke},
  \citenamefont {Maslov}, \citenamefont {Roetteler}, \citenamefont {Debnath},
  \citenamefont {Figgatt}, \citenamefont {Landsman}, \citenamefont {Wright},\
  and\ \citenamefont {Monroe}}]{Linke2017}%
  \BibitemOpen
  \bibfield  {author} {\bibinfo {author} {\bibfnamefont {N.~M.}\ \bibnamefont
  {Linke}}, \bibinfo {author} {\bibfnamefont {D.}~\bibnamefont {Maslov}},
  \bibinfo {author} {\bibfnamefont {M.}~\bibnamefont {Roetteler}}, \bibinfo
  {author} {\bibfnamefont {S.}~\bibnamefont {Debnath}}, \bibinfo {author}
  {\bibfnamefont {C.}~\bibnamefont {Figgatt}}, \bibinfo {author} {\bibfnamefont
  {K.~A.}\ \bibnamefont {Landsman}}, \bibinfo {author} {\bibfnamefont
  {K.}~\bibnamefont {Wright}}, \ and\ \bibinfo {author} {\bibfnamefont
  {C.}~\bibnamefont {Monroe}},\ }\href@noop {} {\bibfield  {journal} {\bibinfo
  {journal} {Proceedings of the National Academy of Sciences}\ ,\ \bibinfo
  {pages} {201618020}} (\bibinfo {year} {2017})}\BibitemShut {NoStop}%
\bibitem [{\citenamefont {Sawyer}\ \emph {et~al.}(2012)\citenamefont {Sawyer},
  \citenamefont {Britton}, \citenamefont {Keith}, \citenamefont {Wang},
  \citenamefont {Freericks}, \citenamefont {Uys}, \citenamefont {Biercuk},\
  and\ \citenamefont {Bollinger}}]{Sawyer2012}%
  \BibitemOpen
  \bibfield  {author} {\bibinfo {author} {\bibfnamefont {B.~C.}\ \bibnamefont
  {Sawyer}}, \bibinfo {author} {\bibfnamefont {J.~W.}\ \bibnamefont {Britton}},
  \bibinfo {author} {\bibfnamefont {A.~C.}\ \bibnamefont {Keith}}, \bibinfo
  {author} {\bibfnamefont {C.-C.~J.}\ \bibnamefont {Wang}}, \bibinfo {author}
  {\bibfnamefont {J.~K.}\ \bibnamefont {Freericks}}, \bibinfo {author}
  {\bibfnamefont {H.}~\bibnamefont {Uys}}, \bibinfo {author} {\bibfnamefont
  {M.~J.}\ \bibnamefont {Biercuk}}, \ and\ \bibinfo {author} {\bibfnamefont
  {J.~J.}\ \bibnamefont {Bollinger}},\ }\href@noop {} {\bibfield  {journal}
  {\bibinfo  {journal} {Physical review letters}\ }\textbf {\bibinfo {volume}
  {108}},\ \bibinfo {pages} {213003} (\bibinfo {year} {2012})}\BibitemShut
  {NoStop}%
\bibitem [{\citenamefont {Yoshimura}\ \emph {et~al.}(2015)\citenamefont
  {Yoshimura}, \citenamefont {Stork}, \citenamefont {Dadic}, \citenamefont
  {Campbell},\ and\ \citenamefont {Freericks}}]{Yoshimura2015}%
  \BibitemOpen
  \bibfield  {author} {\bibinfo {author} {\bibfnamefont {B.}~\bibnamefont
  {Yoshimura}}, \bibinfo {author} {\bibfnamefont {M.}~\bibnamefont {Stork}},
  \bibinfo {author} {\bibfnamefont {D.}~\bibnamefont {Dadic}}, \bibinfo
  {author} {\bibfnamefont {W.~C.}\ \bibnamefont {Campbell}}, \ and\ \bibinfo
  {author} {\bibfnamefont {J.~K.}\ \bibnamefont {Freericks}},\ }\href@noop {}
  {\bibfield  {journal} {\bibinfo  {journal} {EPJ Quantum Technology}\ }\textbf
  {\bibinfo {volume} {2}},\ \bibinfo {pages} {2} (\bibinfo {year}
  {2015})}\BibitemShut {NoStop}%
\bibitem [{\citenamefont {Richerme}(2016)}]{Richerme2016}%
  \BibitemOpen
  \bibfield  {author} {\bibinfo {author} {\bibfnamefont {P.}~\bibnamefont
  {Richerme}},\ }\href@noop {} {\bibfield  {journal} {\bibinfo  {journal}
  {Physical Review A}\ }\textbf {\bibinfo {volume} {94}},\ \bibinfo {pages}
  {032320} (\bibinfo {year} {2016})}\BibitemShut {NoStop}%
\bibitem [{\citenamefont {Li}\ \emph {et~al.}(2017)\citenamefont {Li},
  \citenamefont {Urban}, \citenamefont {Noel}, \citenamefont {Chuang},
  \citenamefont {Xia}, \citenamefont {Ransford}, \citenamefont {Hemmerling},
  \citenamefont {Wang}, \citenamefont {Li}, \citenamefont {H{\"a}ffner} \emph
  {et~al.}}]{Li2017}%
  \BibitemOpen
  \bibfield  {author} {\bibinfo {author} {\bibfnamefont {H.-K.}\ \bibnamefont
  {Li}}, \bibinfo {author} {\bibfnamefont {E.}~\bibnamefont {Urban}}, \bibinfo
  {author} {\bibfnamefont {C.}~\bibnamefont {Noel}}, \bibinfo {author}
  {\bibfnamefont {A.}~\bibnamefont {Chuang}}, \bibinfo {author} {\bibfnamefont
  {Y.}~\bibnamefont {Xia}}, \bibinfo {author} {\bibfnamefont {A.}~\bibnamefont
  {Ransford}}, \bibinfo {author} {\bibfnamefont {B.}~\bibnamefont
  {Hemmerling}}, \bibinfo {author} {\bibfnamefont {Y.}~\bibnamefont {Wang}},
  \bibinfo {author} {\bibfnamefont {T.}~\bibnamefont {Li}}, \bibinfo {author}
  {\bibfnamefont {H.}~\bibnamefont {H{\"a}ffner}},  \emph {et~al.},\
  }\href@noop {} {\bibfield  {journal} {\bibinfo  {journal} {Physical review
  letters}\ }\textbf {\bibinfo {volume} {118}},\ \bibinfo {pages} {053001}
  (\bibinfo {year} {2017})}\BibitemShut {NoStop}%
\bibitem [{\citenamefont {Wineland}\ \emph {et~al.}(1998)\citenamefont
  {Wineland}, \citenamefont {Monroe}, \citenamefont {Itano}, \citenamefont
  {Leibfried}, \citenamefont {King},\ and\ \citenamefont
  {Meekhof}}]{Wineland1998}%
  \BibitemOpen
  \bibfield  {author} {\bibinfo {author} {\bibfnamefont {D.~J.}\ \bibnamefont
  {Wineland}}, \bibinfo {author} {\bibfnamefont {C.}~\bibnamefont {Monroe}},
  \bibinfo {author} {\bibfnamefont {W.~M.}\ \bibnamefont {Itano}}, \bibinfo
  {author} {\bibfnamefont {D.}~\bibnamefont {Leibfried}}, \bibinfo {author}
  {\bibfnamefont {B.~E.}\ \bibnamefont {King}}, \ and\ \bibinfo {author}
  {\bibfnamefont {D.~M.}\ \bibnamefont {Meekhof}},\ }\href@noop {} {\bibfield
  {journal} {\bibinfo  {journal} {Journal of Research of the National Institute
  of Standards and Technology}\ }\textbf {\bibinfo {volume} {103}},\ \bibinfo
  {pages} {259} (\bibinfo {year} {1998})}\BibitemShut {NoStop}%
\bibitem [{\citenamefont {Zhang}\ \emph {et~al.}(2017)\citenamefont {Zhang},
  \citenamefont {Pagano}, \citenamefont {Hess}, \citenamefont {Kyprianidis},
  \citenamefont {Becker}, \citenamefont {Kaplan}, \citenamefont {Gorshkov},
  \citenamefont {Gong},\ and\ \citenamefont {Monroe}}]{Zhang2017}%
  \BibitemOpen
  \bibfield  {author} {\bibinfo {author} {\bibfnamefont {J.}~\bibnamefont
  {Zhang}}, \bibinfo {author} {\bibfnamefont {G.}~\bibnamefont {Pagano}},
  \bibinfo {author} {\bibfnamefont {P.~W.}\ \bibnamefont {Hess}}, \bibinfo
  {author} {\bibfnamefont {A.}~\bibnamefont {Kyprianidis}}, \bibinfo {author}
  {\bibfnamefont {P.}~\bibnamefont {Becker}}, \bibinfo {author} {\bibfnamefont
  {H.}~\bibnamefont {Kaplan}}, \bibinfo {author} {\bibfnamefont {A.~V.}\
  \bibnamefont {Gorshkov}}, \bibinfo {author} {\bibfnamefont {Z.-X.}\
  \bibnamefont {Gong}}, \ and\ \bibinfo {author} {\bibfnamefont
  {C.}~\bibnamefont {Monroe}},\ }\href@noop {} {\bibfield  {journal} {\bibinfo
  {journal} {Nature}\ }\textbf {\bibinfo {volume} {551}},\ \bibinfo {pages}
  {601} (\bibinfo {year} {2017})}\BibitemShut {NoStop}%
\bibitem [{\citenamefont {Pagano}\ \emph {et~al.}(2018)\citenamefont {Pagano},
  \citenamefont {Hess}, \citenamefont {Kaplan}, \citenamefont {Tan},
  \citenamefont {Richerme}, \citenamefont {Becker}, \citenamefont
  {Kyprianidis}, \citenamefont {Zhang}, \citenamefont {Birckelbaw},
  \citenamefont {Hernandez} \emph {et~al.}}]{Pagano2018}%
  \BibitemOpen
  \bibfield  {author} {\bibinfo {author} {\bibfnamefont {G.}~\bibnamefont
  {Pagano}}, \bibinfo {author} {\bibfnamefont {P.}~\bibnamefont {Hess}},
  \bibinfo {author} {\bibfnamefont {H.}~\bibnamefont {Kaplan}}, \bibinfo
  {author} {\bibfnamefont {W.}~\bibnamefont {Tan}}, \bibinfo {author}
  {\bibfnamefont {P.}~\bibnamefont {Richerme}}, \bibinfo {author}
  {\bibfnamefont {P.}~\bibnamefont {Becker}}, \bibinfo {author} {\bibfnamefont
  {A.}~\bibnamefont {Kyprianidis}}, \bibinfo {author} {\bibfnamefont
  {J.}~\bibnamefont {Zhang}}, \bibinfo {author} {\bibfnamefont
  {E.}~\bibnamefont {Birckelbaw}}, \bibinfo {author} {\bibfnamefont
  {M.}~\bibnamefont {Hernandez}},  \emph {et~al.},\ }\href@noop {} {\bibfield
  {journal} {\bibinfo  {journal} {arXiv preprint arXiv:1802.03118}\ } (\bibinfo
  {year} {2018})}\BibitemShut {NoStop}%
\bibitem [{\citenamefont {Friedenauer}\ \emph {et~al.}(2008)\citenamefont
  {Friedenauer}, \citenamefont {Schmitz}, \citenamefont {Glueckert},
  \citenamefont {Porras},\ and\ \citenamefont {Sch{\"a}tz}}]{Friedenauer2008}%
  \BibitemOpen
  \bibfield  {author} {\bibinfo {author} {\bibfnamefont {A.}~\bibnamefont
  {Friedenauer}}, \bibinfo {author} {\bibfnamefont {H.}~\bibnamefont
  {Schmitz}}, \bibinfo {author} {\bibfnamefont {J.~T.}\ \bibnamefont
  {Glueckert}}, \bibinfo {author} {\bibfnamefont {D.}~\bibnamefont {Porras}}, \
  and\ \bibinfo {author} {\bibfnamefont {T.}~\bibnamefont {Sch{\"a}tz}},\
  }\href@noop {} {\bibfield  {journal} {\bibinfo  {journal} {Nature Physics}\
  }\textbf {\bibinfo {volume} {4}},\ \bibinfo {pages} {757} (\bibinfo {year}
  {2008})}\BibitemShut {NoStop}%
\bibitem [{\citenamefont {Gerritsma}\ \emph {et~al.}(2010)\citenamefont
  {Gerritsma}, \citenamefont {Kirchmair}, \citenamefont {Z{\"a}hringer},
  \citenamefont {Solano}, \citenamefont {Blatt},\ and\ \citenamefont
  {Roos}}]{Gerritsma2010}%
  \BibitemOpen
  \bibfield  {author} {\bibinfo {author} {\bibfnamefont {R.}~\bibnamefont
  {Gerritsma}}, \bibinfo {author} {\bibfnamefont {G.}~\bibnamefont
  {Kirchmair}}, \bibinfo {author} {\bibfnamefont {F.}~\bibnamefont
  {Z{\"a}hringer}}, \bibinfo {author} {\bibfnamefont {E.}~\bibnamefont
  {Solano}}, \bibinfo {author} {\bibfnamefont {R.}~\bibnamefont {Blatt}}, \
  and\ \bibinfo {author} {\bibfnamefont {C.}~\bibnamefont {Roos}},\ }\href@noop
  {} {\bibfield  {journal} {\bibinfo  {journal} {Nature}\ }\textbf {\bibinfo
  {volume} {463}},\ \bibinfo {pages} {68} (\bibinfo {year} {2010})}\BibitemShut
  {NoStop}%
\bibitem [{\citenamefont {Gerritsma}\ \emph {et~al.}(2011)\citenamefont
  {Gerritsma}, \citenamefont {Lanyon}, \citenamefont {Kirchmair}, \citenamefont
  {Z{\"a}hringer}, \citenamefont {Hempel}, \citenamefont {Casanova},
  \citenamefont {Garc{\'\i}a-Ripoll}, \citenamefont {Solano}, \citenamefont
  {Blatt},\ and\ \citenamefont {Roos}}]{Gerritsma2011}%
  \BibitemOpen
  \bibfield  {author} {\bibinfo {author} {\bibfnamefont {R.}~\bibnamefont
  {Gerritsma}}, \bibinfo {author} {\bibfnamefont {B.}~\bibnamefont {Lanyon}},
  \bibinfo {author} {\bibfnamefont {G.}~\bibnamefont {Kirchmair}}, \bibinfo
  {author} {\bibfnamefont {F.}~\bibnamefont {Z{\"a}hringer}}, \bibinfo {author}
  {\bibfnamefont {C.}~\bibnamefont {Hempel}}, \bibinfo {author} {\bibfnamefont
  {J.}~\bibnamefont {Casanova}}, \bibinfo {author} {\bibfnamefont {J.~J.}\
  \bibnamefont {Garc{\'\i}a-Ripoll}}, \bibinfo {author} {\bibfnamefont
  {E.}~\bibnamefont {Solano}}, \bibinfo {author} {\bibfnamefont
  {R.}~\bibnamefont {Blatt}}, \ and\ \bibinfo {author} {\bibfnamefont {C.~F.}\
  \bibnamefont {Roos}},\ }\href@noop {} {\bibfield  {journal} {\bibinfo
  {journal} {Physical review letters}\ }\textbf {\bibinfo {volume} {106}},\
  \bibinfo {pages} {060503} (\bibinfo {year} {2011})}\BibitemShut {NoStop}%
\bibitem [{\citenamefont {Islam}\ \emph {et~al.}(2011)\citenamefont {Islam},
  \citenamefont {Edwards}, \citenamefont {Kim}, \citenamefont {Korenblit},
  \citenamefont {Noh}, \citenamefont {Carmichael}, \citenamefont {Lin},
  \citenamefont {Duan}, \citenamefont {Wang}, \citenamefont {Freericks} \emph
  {et~al.}}]{Islam2011}%
  \BibitemOpen
  \bibfield  {author} {\bibinfo {author} {\bibfnamefont {R.}~\bibnamefont
  {Islam}}, \bibinfo {author} {\bibfnamefont {E.}~\bibnamefont {Edwards}},
  \bibinfo {author} {\bibfnamefont {K.}~\bibnamefont {Kim}}, \bibinfo {author}
  {\bibfnamefont {S.}~\bibnamefont {Korenblit}}, \bibinfo {author}
  {\bibfnamefont {C.}~\bibnamefont {Noh}}, \bibinfo {author} {\bibfnamefont
  {H.}~\bibnamefont {Carmichael}}, \bibinfo {author} {\bibfnamefont {G.-D.}\
  \bibnamefont {Lin}}, \bibinfo {author} {\bibfnamefont {L.-M.}\ \bibnamefont
  {Duan}}, \bibinfo {author} {\bibfnamefont {C.-C.~J.}\ \bibnamefont {Wang}},
  \bibinfo {author} {\bibfnamefont {J.}~\bibnamefont {Freericks}},  \emph
  {et~al.},\ }\href@noop {} {\bibfield  {journal} {\bibinfo  {journal} {Nature
  communications}\ }\textbf {\bibinfo {volume} {2}},\ \bibinfo {pages} {377}
  (\bibinfo {year} {2011})}\BibitemShut {NoStop}%
\bibitem [{\citenamefont {Senko}\ \emph {et~al.}(2015)\citenamefont {Senko},
  \citenamefont {Richerme}, \citenamefont {Smith}, \citenamefont {Lee},
  \citenamefont {Cohen}, \citenamefont {Retzker},\ and\ \citenamefont
  {Monroe}}]{Senko2015}%
  \BibitemOpen
  \bibfield  {author} {\bibinfo {author} {\bibfnamefont {C.}~\bibnamefont
  {Senko}}, \bibinfo {author} {\bibfnamefont {P.}~\bibnamefont {Richerme}},
  \bibinfo {author} {\bibfnamefont {J.}~\bibnamefont {Smith}}, \bibinfo
  {author} {\bibfnamefont {A.}~\bibnamefont {Lee}}, \bibinfo {author}
  {\bibfnamefont {I.}~\bibnamefont {Cohen}}, \bibinfo {author} {\bibfnamefont
  {A.}~\bibnamefont {Retzker}}, \ and\ \bibinfo {author} {\bibfnamefont
  {C.}~\bibnamefont {Monroe}},\ }\href@noop {} {\bibfield  {journal} {\bibinfo
  {journal} {Physical Review X}\ }\textbf {\bibinfo {volume} {5}},\ \bibinfo
  {pages} {021026} (\bibinfo {year} {2015})}\BibitemShut {NoStop}%
\bibitem [{\citenamefont {Jurcevic}\ \emph {et~al.}(2017)\citenamefont
  {Jurcevic}, \citenamefont {Shen}, \citenamefont {Hauke}, \citenamefont
  {Maier}, \citenamefont {Brydges}, \citenamefont {Hempel}, \citenamefont
  {Lanyon}, \citenamefont {Heyl}, \citenamefont {Blatt},\ and\ \citenamefont
  {Roos}}]{Jurcevic2017}%
  \BibitemOpen
  \bibfield  {author} {\bibinfo {author} {\bibfnamefont {P.}~\bibnamefont
  {Jurcevic}}, \bibinfo {author} {\bibfnamefont {H.}~\bibnamefont {Shen}},
  \bibinfo {author} {\bibfnamefont {P.}~\bibnamefont {Hauke}}, \bibinfo
  {author} {\bibfnamefont {C.}~\bibnamefont {Maier}}, \bibinfo {author}
  {\bibfnamefont {T.}~\bibnamefont {Brydges}}, \bibinfo {author} {\bibfnamefont
  {C.}~\bibnamefont {Hempel}}, \bibinfo {author} {\bibfnamefont
  {B.}~\bibnamefont {Lanyon}}, \bibinfo {author} {\bibfnamefont
  {M.}~\bibnamefont {Heyl}}, \bibinfo {author} {\bibfnamefont {R.}~\bibnamefont
  {Blatt}}, \ and\ \bibinfo {author} {\bibfnamefont {C.}~\bibnamefont {Roos}},\
  }\href@noop {} {\bibfield  {journal} {\bibinfo  {journal} {Physical review
  letters}\ }\textbf {\bibinfo {volume} {119}},\ \bibinfo {pages} {080501}
  (\bibinfo {year} {2017})}\BibitemShut {NoStop}%
\bibitem [{\citenamefont {Lanyon}\ \emph {et~al.}(2011)\citenamefont {Lanyon},
  \citenamefont {Hempel}, \citenamefont {Nigg}, \citenamefont {M{\"u}ller},
  \citenamefont {Gerritsma}, \citenamefont {Z{\"a}hringer}, \citenamefont
  {Schindler}, \citenamefont {Barreiro}, \citenamefont {Rambach}, \citenamefont
  {Kirchmair} \emph {et~al.}}]{Lanyon2011}%
  \BibitemOpen
  \bibfield  {author} {\bibinfo {author} {\bibfnamefont {B.~P.}\ \bibnamefont
  {Lanyon}}, \bibinfo {author} {\bibfnamefont {C.}~\bibnamefont {Hempel}},
  \bibinfo {author} {\bibfnamefont {D.}~\bibnamefont {Nigg}}, \bibinfo {author}
  {\bibfnamefont {M.}~\bibnamefont {M{\"u}ller}}, \bibinfo {author}
  {\bibfnamefont {R.}~\bibnamefont {Gerritsma}}, \bibinfo {author}
  {\bibfnamefont {F.}~\bibnamefont {Z{\"a}hringer}}, \bibinfo {author}
  {\bibfnamefont {P.}~\bibnamefont {Schindler}}, \bibinfo {author}
  {\bibfnamefont {J.}~\bibnamefont {Barreiro}}, \bibinfo {author}
  {\bibfnamefont {M.}~\bibnamefont {Rambach}}, \bibinfo {author} {\bibfnamefont
  {G.}~\bibnamefont {Kirchmair}},  \emph {et~al.},\ }\href@noop {} {\bibfield
  {journal} {\bibinfo  {journal} {Science}\ }\textbf {\bibinfo {volume}
  {334}},\ \bibinfo {pages} {57} (\bibinfo {year} {2011})}\BibitemShut
  {NoStop}%
\bibitem [{\citenamefont {Barreiro}\ \emph {et~al.}(2011)\citenamefont
  {Barreiro}, \citenamefont {M{\"u}ller}, \citenamefont {Schindler},
  \citenamefont {Nigg}, \citenamefont {Monz}, \citenamefont {Chwalla},
  \citenamefont {Hennrich}, \citenamefont {Roos}, \citenamefont {Zoller},\ and\
  \citenamefont {Blatt}}]{Barreiro2011}%
  \BibitemOpen
  \bibfield  {author} {\bibinfo {author} {\bibfnamefont {J.~T.}\ \bibnamefont
  {Barreiro}}, \bibinfo {author} {\bibfnamefont {M.}~\bibnamefont
  {M{\"u}ller}}, \bibinfo {author} {\bibfnamefont {P.}~\bibnamefont
  {Schindler}}, \bibinfo {author} {\bibfnamefont {D.}~\bibnamefont {Nigg}},
  \bibinfo {author} {\bibfnamefont {T.}~\bibnamefont {Monz}}, \bibinfo {author}
  {\bibfnamefont {M.}~\bibnamefont {Chwalla}}, \bibinfo {author} {\bibfnamefont
  {M.}~\bibnamefont {Hennrich}}, \bibinfo {author} {\bibfnamefont {C.~F.}\
  \bibnamefont {Roos}}, \bibinfo {author} {\bibfnamefont {P.}~\bibnamefont
  {Zoller}}, \ and\ \bibinfo {author} {\bibfnamefont {R.}~\bibnamefont
  {Blatt}},\ }\href@noop {} {\bibfield  {journal} {\bibinfo  {journal}
  {Nature}\ }\textbf {\bibinfo {volume} {470}},\ \bibinfo {pages} {486}
  (\bibinfo {year} {2011})}\BibitemShut {NoStop}%
\bibitem [{\citenamefont {{Hempel}}\ \emph {et~al.}(2018)\citenamefont
  {{Hempel}}, \citenamefont {{Maier}}, \citenamefont {{Romero}}, \citenamefont
  {{McClean}}, \citenamefont {{Monz}}, \citenamefont {{Shen}}, \citenamefont
  {{Jurcevic}}, \citenamefont {{Lanyon}}, \citenamefont {{Love}}, \citenamefont
  {{Babbush}}, \citenamefont {{Aspuru-Guzik}}, \citenamefont {{Blatt}},\ and\
  \citenamefont {{Roos}}}]{Hempel2018}%
  \BibitemOpen
  \bibfield  {author} {\bibinfo {author} {\bibfnamefont {C.}~\bibnamefont
  {{Hempel}}}, \bibinfo {author} {\bibfnamefont {C.}~\bibnamefont {{Maier}}},
  \bibinfo {author} {\bibfnamefont {J.}~\bibnamefont {{Romero}}}, \bibinfo
  {author} {\bibfnamefont {J.}~\bibnamefont {{McClean}}}, \bibinfo {author}
  {\bibfnamefont {T.}~\bibnamefont {{Monz}}}, \bibinfo {author} {\bibfnamefont
  {H.}~\bibnamefont {{Shen}}}, \bibinfo {author} {\bibfnamefont
  {P.}~\bibnamefont {{Jurcevic}}}, \bibinfo {author} {\bibfnamefont {B.~P.}\
  \bibnamefont {{Lanyon}}}, \bibinfo {author} {\bibfnamefont {P.}~\bibnamefont
  {{Love}}}, \bibinfo {author} {\bibfnamefont {R.}~\bibnamefont {{Babbush}}},
  \bibinfo {author} {\bibfnamefont {A.}~\bibnamefont {{Aspuru-Guzik}}},
  \bibinfo {author} {\bibfnamefont {R.}~\bibnamefont {{Blatt}}}, \ and\
  \bibinfo {author} {\bibfnamefont {C.~F.}\ \bibnamefont {{Roos}}},\ }\href
  {\doibase 10.1103/PhysRevX.8.031022} {\bibfield  {journal} {\bibinfo
  {journal} {Physical Review X}\ }\textbf {\bibinfo {volume} {8}},\ \bibinfo
  {eid} {031022} (\bibinfo {year} {2018})},\ \Eprint
  {http://arxiv.org/abs/1803.10238} {arXiv:1803.10238 [quant-ph]} \BibitemShut
  {NoStop}%
\bibitem [{\citenamefont {Warren}\ \emph {et~al.}(1979)\citenamefont {Warren},
  \citenamefont {Sinton}, \citenamefont {Weitekamp},\ and\ \citenamefont
  {Pines}}]{Warren1979}%
  \BibitemOpen
  \bibfield  {author} {\bibinfo {author} {\bibfnamefont {W.}~\bibnamefont
  {Warren}}, \bibinfo {author} {\bibfnamefont {S.}~\bibnamefont {Sinton}},
  \bibinfo {author} {\bibfnamefont {D.}~\bibnamefont {Weitekamp}}, \ and\
  \bibinfo {author} {\bibfnamefont {A.}~\bibnamefont {Pines}},\ }\href@noop {}
  {\bibfield  {journal} {\bibinfo  {journal} {Physical Review Letters}\
  }\textbf {\bibinfo {volume} {43}},\ \bibinfo {pages} {1791} (\bibinfo {year}
  {1979})}\BibitemShut {NoStop}%
\bibitem [{\citenamefont {Baum}\ \emph {et~al.}(1985)\citenamefont {Baum},
  \citenamefont {Munowitz}, \citenamefont {Garroway},\ and\ \citenamefont
  {Pines}}]{Baum1985}%
  \BibitemOpen
  \bibfield  {author} {\bibinfo {author} {\bibfnamefont {J.}~\bibnamefont
  {Baum}}, \bibinfo {author} {\bibfnamefont {M.}~\bibnamefont {Munowitz}},
  \bibinfo {author} {\bibfnamefont {A.}~\bibnamefont {Garroway}}, \ and\
  \bibinfo {author} {\bibfnamefont {A.}~\bibnamefont {Pines}},\ }\href@noop {}
  {\bibfield  {journal} {\bibinfo  {journal} {The Journal of chemical physics}\
  }\textbf {\bibinfo {volume} {83}},\ \bibinfo {pages} {2015} (\bibinfo {year}
  {1985})}\BibitemShut {NoStop}%
\bibitem [{\citenamefont {Ajoy}\ and\ \citenamefont
  {Cappellaro}(2013)}]{Ajoy2013}%
  \BibitemOpen
  \bibfield  {author} {\bibinfo {author} {\bibfnamefont {A.}~\bibnamefont
  {Ajoy}}\ and\ \bibinfo {author} {\bibfnamefont {P.}~\bibnamefont
  {Cappellaro}},\ }\href@noop {} {\bibfield  {journal} {\bibinfo  {journal}
  {Physical review letters}\ }\textbf {\bibinfo {volume} {110}},\ \bibinfo
  {pages} {220503} (\bibinfo {year} {2013})}\BibitemShut {NoStop}%
\bibitem [{\citenamefont {Heyl}\ \emph {et~al.}(2013)\citenamefont {Heyl},
  \citenamefont {Polkovnikov},\ and\ \citenamefont {Kehrein}}]{Heyl2013}%
  \BibitemOpen
  \bibfield  {author} {\bibinfo {author} {\bibfnamefont {M.}~\bibnamefont
  {Heyl}}, \bibinfo {author} {\bibfnamefont {A.}~\bibnamefont {Polkovnikov}}, \
  and\ \bibinfo {author} {\bibfnamefont {S.}~\bibnamefont {Kehrein}},\
  }\href@noop {} {\bibfield  {journal} {\bibinfo  {journal} {Physical review
  letters}\ }\textbf {\bibinfo {volume} {110}},\ \bibinfo {pages} {135704}
  (\bibinfo {year} {2013})}\BibitemShut {NoStop}%
\bibitem [{\citenamefont {Vosk}\ and\ \citenamefont {Altman}(2014)}]{Vosk2014}%
  \BibitemOpen
  \bibfield  {author} {\bibinfo {author} {\bibfnamefont {R.}~\bibnamefont
  {Vosk}}\ and\ \bibinfo {author} {\bibfnamefont {E.}~\bibnamefont {Altman}},\
  }\href@noop {} {\bibfield  {journal} {\bibinfo  {journal} {Physical review
  letters}\ }\textbf {\bibinfo {volume} {112}},\ \bibinfo {pages} {217204}
  (\bibinfo {year} {2014})}\BibitemShut {NoStop}%
\bibitem [{\citenamefont {Rigol}\ \emph {et~al.}(2008)\citenamefont {Rigol},
  \citenamefont {Dunjko},\ and\ \citenamefont {Olshanii}}]{Rigol2008}%
  \BibitemOpen
  \bibfield  {author} {\bibinfo {author} {\bibfnamefont {M.}~\bibnamefont
  {Rigol}}, \bibinfo {author} {\bibfnamefont {V.}~\bibnamefont {Dunjko}}, \
  and\ \bibinfo {author} {\bibfnamefont {M.}~\bibnamefont {Olshanii}},\
  }\href@noop {} {\bibfield  {journal} {\bibinfo  {journal} {Nature}\ }\textbf
  {\bibinfo {volume} {452}},\ \bibinfo {pages} {854} (\bibinfo {year}
  {2008})}\BibitemShut {NoStop}%
\bibitem [{\citenamefont {{Schreiber}}\ \emph {et~al.}(2015)\citenamefont
  {{Schreiber}}, \citenamefont {{Hodgman}}, \citenamefont {{Bordia}},
  \citenamefont {{L{\"u}schen}}, \citenamefont {{Fischer}}, \citenamefont
  {{Vosk}}, \citenamefont {{Altman}}, \citenamefont {{Schneider}},\ and\
  \citenamefont {{Bloch}}}]{Schreiber2015}%
  \BibitemOpen
  \bibfield  {author} {\bibinfo {author} {\bibfnamefont {M.}~\bibnamefont
  {{Schreiber}}}, \bibinfo {author} {\bibfnamefont {S.~S.}\ \bibnamefont
  {{Hodgman}}}, \bibinfo {author} {\bibfnamefont {P.}~\bibnamefont {{Bordia}}},
  \bibinfo {author} {\bibfnamefont {H.~P.}\ \bibnamefont {{L{\"u}schen}}},
  \bibinfo {author} {\bibfnamefont {M.~H.}\ \bibnamefont {{Fischer}}}, \bibinfo
  {author} {\bibfnamefont {R.}~\bibnamefont {{Vosk}}}, \bibinfo {author}
  {\bibfnamefont {E.}~\bibnamefont {{Altman}}}, \bibinfo {author}
  {\bibfnamefont {U.}~\bibnamefont {{Schneider}}}, \ and\ \bibinfo {author}
  {\bibfnamefont {I.}~\bibnamefont {{Bloch}}},\ }\href {\doibase
  10.1126/science.aaa7432} {\bibfield  {journal} {\bibinfo  {journal}
  {Science}\ }\textbf {\bibinfo {volume} {349}},\ \bibinfo {pages} {842}
  (\bibinfo {year} {2015})},\ \Eprint {http://arxiv.org/abs/1501.05661}
  {arXiv:1501.05661 [cond-mat.quant-gas]} \BibitemShut {NoStop}%
\bibitem [{\citenamefont {{Smith}}\ \emph {et~al.}(2016)\citenamefont
  {{Smith}}, \citenamefont {{Lee}}, \citenamefont {{Richerme}}, \citenamefont
  {{Neyenhuis}}, \citenamefont {{Hess}}, \citenamefont {{Hauke}}, \citenamefont
  {{Heyl}}, \citenamefont {{Huse}},\ and\ \citenamefont
  {{Monroe}}}]{Smith2016}%
  \BibitemOpen
  \bibfield  {author} {\bibinfo {author} {\bibfnamefont {J.}~\bibnamefont
  {{Smith}}}, \bibinfo {author} {\bibfnamefont {A.}~\bibnamefont {{Lee}}},
  \bibinfo {author} {\bibfnamefont {P.}~\bibnamefont {{Richerme}}}, \bibinfo
  {author} {\bibfnamefont {B.}~\bibnamefont {{Neyenhuis}}}, \bibinfo {author}
  {\bibfnamefont {P.~W.}\ \bibnamefont {{Hess}}}, \bibinfo {author}
  {\bibfnamefont {P.}~\bibnamefont {{Hauke}}}, \bibinfo {author} {\bibfnamefont
  {M.}~\bibnamefont {{Heyl}}}, \bibinfo {author} {\bibfnamefont {D.~A.}\
  \bibnamefont {{Huse}}}, \ and\ \bibinfo {author} {\bibfnamefont
  {C.}~\bibnamefont {{Monroe}}},\ }\href {\doibase 10.1038/nphys3783}
  {\bibfield  {journal} {\bibinfo  {journal} {Nature Physics}\ }\textbf
  {\bibinfo {volume} {12}},\ \bibinfo {pages} {907} (\bibinfo {year} {2016})},\
  \Eprint {http://arxiv.org/abs/1508.07026} {arXiv:1508.07026 [quant-ph]}
  \BibitemShut {NoStop}%
\bibitem [{\citenamefont {Bordia}\ \emph {et~al.}(2016)\citenamefont {Bordia},
  \citenamefont {L{\"u}schen}, \citenamefont {Hodgman}, \citenamefont
  {Schreiber}, \citenamefont {Bloch},\ and\ \citenamefont
  {Schneider}}]{Bordia2016}%
  \BibitemOpen
  \bibfield  {author} {\bibinfo {author} {\bibfnamefont {P.}~\bibnamefont
  {Bordia}}, \bibinfo {author} {\bibfnamefont {H.~P.}\ \bibnamefont
  {L{\"u}schen}}, \bibinfo {author} {\bibfnamefont {S.~S.}\ \bibnamefont
  {Hodgman}}, \bibinfo {author} {\bibfnamefont {M.}~\bibnamefont {Schreiber}},
  \bibinfo {author} {\bibfnamefont {I.}~\bibnamefont {Bloch}}, \ and\ \bibinfo
  {author} {\bibfnamefont {U.}~\bibnamefont {Schneider}},\ }\href@noop {}
  {\bibfield  {journal} {\bibinfo  {journal} {Physical review letters}\
  }\textbf {\bibinfo {volume} {116}},\ \bibinfo {pages} {140401} (\bibinfo
  {year} {2016})}\BibitemShut {NoStop}%
\bibitem [{\citenamefont {Choi}\ \emph {et~al.}(2016)\citenamefont {Choi},
  \citenamefont {Hild}, \citenamefont {Zeiher}, \citenamefont {Schau{\ss}},
  \citenamefont {Rubio-Abadal}, \citenamefont {Yefsah}, \citenamefont
  {Khemani}, \citenamefont {Huse}, \citenamefont {Bloch},\ and\ \citenamefont
  {Gross}}]{Choi2016}%
  \BibitemOpen
  \bibfield  {author} {\bibinfo {author} {\bibfnamefont {J.-y.}\ \bibnamefont
  {Choi}}, \bibinfo {author} {\bibfnamefont {S.}~\bibnamefont {Hild}}, \bibinfo
  {author} {\bibfnamefont {J.}~\bibnamefont {Zeiher}}, \bibinfo {author}
  {\bibfnamefont {P.}~\bibnamefont {Schau{\ss}}}, \bibinfo {author}
  {\bibfnamefont {A.}~\bibnamefont {Rubio-Abadal}}, \bibinfo {author}
  {\bibfnamefont {T.}~\bibnamefont {Yefsah}}, \bibinfo {author} {\bibfnamefont
  {V.}~\bibnamefont {Khemani}}, \bibinfo {author} {\bibfnamefont {D.~A.}\
  \bibnamefont {Huse}}, \bibinfo {author} {\bibfnamefont {I.}~\bibnamefont
  {Bloch}}, \ and\ \bibinfo {author} {\bibfnamefont {C.}~\bibnamefont
  {Gross}},\ }\href@noop {} {\bibfield  {journal} {\bibinfo  {journal}
  {Science}\ }\textbf {\bibinfo {volume} {352}},\ \bibinfo {pages} {1547}
  (\bibinfo {year} {2016})}\BibitemShut {NoStop}%
\bibitem [{\citenamefont {L{\"u}schen}\ \emph {et~al.}(2017)\citenamefont
  {L{\"u}schen}, \citenamefont {Bordia}, \citenamefont {Hodgman}, \citenamefont
  {Schreiber}, \citenamefont {Sarkar}, \citenamefont {Daley}, \citenamefont
  {Fischer}, \citenamefont {Altman}, \citenamefont {Bloch},\ and\ \citenamefont
  {Schneider}}]{Luschen2017}%
  \BibitemOpen
  \bibfield  {author} {\bibinfo {author} {\bibfnamefont {H.~P.}\ \bibnamefont
  {L{\"u}schen}}, \bibinfo {author} {\bibfnamefont {P.}~\bibnamefont {Bordia}},
  \bibinfo {author} {\bibfnamefont {S.~S.}\ \bibnamefont {Hodgman}}, \bibinfo
  {author} {\bibfnamefont {M.}~\bibnamefont {Schreiber}}, \bibinfo {author}
  {\bibfnamefont {S.}~\bibnamefont {Sarkar}}, \bibinfo {author} {\bibfnamefont
  {A.~J.}\ \bibnamefont {Daley}}, \bibinfo {author} {\bibfnamefont {M.~H.}\
  \bibnamefont {Fischer}}, \bibinfo {author} {\bibfnamefont {E.}~\bibnamefont
  {Altman}}, \bibinfo {author} {\bibfnamefont {I.}~\bibnamefont {Bloch}}, \
  and\ \bibinfo {author} {\bibfnamefont {U.}~\bibnamefont {Schneider}},\
  }\href@noop {} {\bibfield  {journal} {\bibinfo  {journal} {Physical Review
  X}\ }\textbf {\bibinfo {volume} {7}},\ \bibinfo {pages} {011034} (\bibinfo
  {year} {2017})}\BibitemShut {NoStop}%
\bibitem [{\citenamefont {Wang}\ \emph {et~al.}(2017)\citenamefont {Wang},
  \citenamefont {Um}, \citenamefont {Zhang}, \citenamefont {An}, \citenamefont
  {Lyu}, \citenamefont {Zhang}, \citenamefont {Duan}, \citenamefont {Yum},\
  and\ \citenamefont {Kim}}]{Wang2017}%
  \BibitemOpen
  \bibfield  {author} {\bibinfo {author} {\bibfnamefont {Y.}~\bibnamefont
  {Wang}}, \bibinfo {author} {\bibfnamefont {M.}~\bibnamefont {Um}}, \bibinfo
  {author} {\bibfnamefont {J.}~\bibnamefont {Zhang}}, \bibinfo {author}
  {\bibfnamefont {S.}~\bibnamefont {An}}, \bibinfo {author} {\bibfnamefont
  {M.}~\bibnamefont {Lyu}}, \bibinfo {author} {\bibfnamefont {J.-N.}\
  \bibnamefont {Zhang}}, \bibinfo {author} {\bibfnamefont {L.-M.}\ \bibnamefont
  {Duan}}, \bibinfo {author} {\bibfnamefont {D.}~\bibnamefont {Yum}}, \ and\
  \bibinfo {author} {\bibfnamefont {K.}~\bibnamefont {Kim}},\ }\href@noop {}
  {\bibfield  {journal} {\bibinfo  {journal} {Nature Photonics}\ }\textbf
  {\bibinfo {volume} {11}},\ \bibinfo {pages} {646} (\bibinfo {year}
  {2017})}\BibitemShut {NoStop}%
\bibitem [{\citenamefont {Ospelkaus}\ \emph {et~al.}(2011)\citenamefont
  {Ospelkaus}, \citenamefont {Warring}, \citenamefont {Colombe}, \citenamefont
  {Brown}, \citenamefont {Amini}, \citenamefont {Leibfried},\ and\
  \citenamefont {Wineland}}]{Ospelkaus2011}%
  \BibitemOpen
  \bibfield  {author} {\bibinfo {author} {\bibfnamefont {C.}~\bibnamefont
  {Ospelkaus}}, \bibinfo {author} {\bibfnamefont {U.}~\bibnamefont {Warring}},
  \bibinfo {author} {\bibfnamefont {Y.}~\bibnamefont {Colombe}}, \bibinfo
  {author} {\bibfnamefont {K.}~\bibnamefont {Brown}}, \bibinfo {author}
  {\bibfnamefont {J.}~\bibnamefont {Amini}}, \bibinfo {author} {\bibfnamefont
  {D.}~\bibnamefont {Leibfried}}, \ and\ \bibinfo {author} {\bibfnamefont
  {D.}~\bibnamefont {Wineland}},\ }\href@noop {} {\bibfield  {journal}
  {\bibinfo  {journal} {Nature}\ }\textbf {\bibinfo {volume} {476}},\ \bibinfo
  {pages} {181} (\bibinfo {year} {2011})}\BibitemShut {NoStop}%
\bibitem [{\citenamefont {Cohen}\ \emph {et~al.}(2015)\citenamefont {Cohen},
  \citenamefont {Weidt}, \citenamefont {Hensinger},\ and\ \citenamefont
  {Retzker}}]{Cohen2015}%
  \BibitemOpen
  \bibfield  {author} {\bibinfo {author} {\bibfnamefont {I.}~\bibnamefont
  {Cohen}}, \bibinfo {author} {\bibfnamefont {S.}~\bibnamefont {Weidt}},
  \bibinfo {author} {\bibfnamefont {W.~K.}\ \bibnamefont {Hensinger}}, \ and\
  \bibinfo {author} {\bibfnamefont {A.}~\bibnamefont {Retzker}},\ }\href@noop
  {} {\bibfield  {journal} {\bibinfo  {journal} {New Journal of Physics}\
  }\textbf {\bibinfo {volume} {17}},\ \bibinfo {pages} {043008} (\bibinfo
  {year} {2015})}\BibitemShut {NoStop}%
\bibitem [{\citenamefont {Cirac}\ and\ \citenamefont
  {Zoller}(1995)}]{Cirac1995}%
  \BibitemOpen
  \bibfield  {author} {\bibinfo {author} {\bibfnamefont {J.~I.}\ \bibnamefont
  {Cirac}}\ and\ \bibinfo {author} {\bibfnamefont {P.}~\bibnamefont {Zoller}},\
  }\href@noop {} {\bibfield  {journal} {\bibinfo  {journal} {Physical review
  letters}\ }\textbf {\bibinfo {volume} {74}},\ \bibinfo {pages} {4091}
  (\bibinfo {year} {1995})}\BibitemShut {NoStop}%
\bibitem [{\citenamefont {Milburn}\ \emph {et~al.}(2000)\citenamefont
  {Milburn}, \citenamefont {Schneider},\ and\ \citenamefont
  {James}}]{Milburn2000}%
  \BibitemOpen
  \bibfield  {author} {\bibinfo {author} {\bibfnamefont {G.}~\bibnamefont
  {Milburn}}, \bibinfo {author} {\bibfnamefont {S.}~\bibnamefont {Schneider}},
  \ and\ \bibinfo {author} {\bibfnamefont {D.}~\bibnamefont {James}},\
  }\href@noop {} {\bibfield  {journal} {\bibinfo  {journal} {Fortschritte der
  Physik}\ }\textbf {\bibinfo {volume} {48}},\ \bibinfo {pages} {801} (\bibinfo
  {year} {2000})}\BibitemShut {NoStop}%
\bibitem [{\citenamefont {Leibfried}\ \emph {et~al.}(2003)\citenamefont
  {Leibfried}, \citenamefont {Blatt}, \citenamefont {Monroe},\ and\
  \citenamefont {Wineland}}]{Leibfried2003}%
  \BibitemOpen
  \bibfield  {author} {\bibinfo {author} {\bibfnamefont {D.}~\bibnamefont
  {Leibfried}}, \bibinfo {author} {\bibfnamefont {R.}~\bibnamefont {Blatt}},
  \bibinfo {author} {\bibfnamefont {C.}~\bibnamefont {Monroe}}, \ and\ \bibinfo
  {author} {\bibfnamefont {D.}~\bibnamefont {Wineland}},\ }\href@noop {}
  {\bibfield  {journal} {\bibinfo  {journal} {Reviews of Modern Physics}\
  }\textbf {\bibinfo {volume} {75}},\ \bibinfo {pages} {281} (\bibinfo {year}
  {2003})}\BibitemShut {NoStop}%
\bibitem [{\citenamefont {Porras}\ and\ \citenamefont
  {Cirac}(2004)}]{Porras2004}%
  \BibitemOpen
  \bibfield  {author} {\bibinfo {author} {\bibfnamefont {D.}~\bibnamefont
  {Porras}}\ and\ \bibinfo {author} {\bibfnamefont {J.~I.}\ \bibnamefont
  {Cirac}},\ }\href@noop {} {\bibfield  {journal} {\bibinfo  {journal}
  {Physical review letters}\ }\textbf {\bibinfo {volume} {92}},\ \bibinfo
  {pages} {207901} (\bibinfo {year} {2004})}\BibitemShut {NoStop}%
\bibitem [{\citenamefont {Haeberlen}\ and\ \citenamefont
  {Waugh}(1968)}]{Haeberlen1968}%
  \BibitemOpen
  \bibfield  {author} {\bibinfo {author} {\bibfnamefont {U.}~\bibnamefont
  {Haeberlen}}\ and\ \bibinfo {author} {\bibfnamefont {J.}~\bibnamefont
  {Waugh}},\ }\href@noop {} {\bibfield  {journal} {\bibinfo  {journal}
  {Physical Review}\ }\textbf {\bibinfo {volume} {175}},\ \bibinfo {pages}
  {453} (\bibinfo {year} {1968})}\BibitemShut {NoStop}%
\bibitem [{\citenamefont {Maricq}(1982)}]{Maricq1982}%
  \BibitemOpen
  \bibfield  {author} {\bibinfo {author} {\bibfnamefont {M.~M.}\ \bibnamefont
  {Maricq}},\ }\href@noop {} {\bibfield  {journal} {\bibinfo  {journal}
  {Physical Review B}\ }\textbf {\bibinfo {volume} {25}},\ \bibinfo {pages}
  {6622} (\bibinfo {year} {1982})}\BibitemShut {NoStop}%
\bibitem [{\citenamefont {Johansson}\ \emph {et~al.}(2012)\citenamefont
  {Johansson}, \citenamefont {Nation},\ and\ \citenamefont
  {Nori}}]{Johansson2012}%
  \BibitemOpen
  \bibfield  {author} {\bibinfo {author} {\bibfnamefont {J.}~\bibnamefont
  {Johansson}}, \bibinfo {author} {\bibfnamefont {P.}~\bibnamefont {Nation}}, \
  and\ \bibinfo {author} {\bibfnamefont {F.}~\bibnamefont {Nori}},\ }\href@noop
  {} {\bibfield  {journal} {\bibinfo  {journal} {Computer Physics
  Communications}\ }\textbf {\bibinfo {volume} {183}},\ \bibinfo {pages} {1760}
  (\bibinfo {year} {2012})}\BibitemShut {NoStop}%
\bibitem [{\citenamefont {{Johansson}}\ \emph {et~al.}(2013)\citenamefont
  {{Johansson}}, \citenamefont {{Nation}},\ and\ \citenamefont
  {{Nori}}}]{Johansson2013}%
  \BibitemOpen
  \bibfield  {author} {\bibinfo {author} {\bibfnamefont {J.~R.}\ \bibnamefont
  {{Johansson}}}, \bibinfo {author} {\bibfnamefont {P.~D.}\ \bibnamefont
  {{Nation}}}, \ and\ \bibinfo {author} {\bibfnamefont {F.}~\bibnamefont
  {{Nori}}},\ }\href {\doibase 10.1016/j.cpc.2012.11.019} {\bibfield  {journal}
  {\bibinfo  {journal} {Computer Physics Communications}\ }\textbf {\bibinfo
  {volume} {184}},\ \bibinfo {pages} {1234} (\bibinfo {year} {2013})},\ \Eprint
  {http://arxiv.org/abs/1211.6518} {arXiv:1211.6518 [quant-ph]} \BibitemShut
  {NoStop}%
\bibitem [{\citenamefont {Schiffer}(1993)}]{Schiffer1993}%
  \BibitemOpen
  \bibfield  {author} {\bibinfo {author} {\bibfnamefont {J.}~\bibnamefont
  {Schiffer}},\ }\href@noop {} {\bibfield  {journal} {\bibinfo  {journal}
  {Physical review letters}\ }\textbf {\bibinfo {volume} {70}},\ \bibinfo
  {pages} {818} (\bibinfo {year} {1993})}\BibitemShut {NoStop}%
\bibitem [{\citenamefont {Olmschenk}\ \emph {et~al.}(2007)\citenamefont
  {Olmschenk}, \citenamefont {Younge}, \citenamefont {Moehring}, \citenamefont
  {Matsukevich}, \citenamefont {Maunz},\ and\ \citenamefont
  {Monroe}}]{Olmschenk2007}%
  \BibitemOpen
  \bibfield  {author} {\bibinfo {author} {\bibfnamefont {S.}~\bibnamefont
  {Olmschenk}}, \bibinfo {author} {\bibfnamefont {K.}~\bibnamefont {Younge}},
  \bibinfo {author} {\bibfnamefont {D.}~\bibnamefont {Moehring}}, \bibinfo
  {author} {\bibfnamefont {D.}~\bibnamefont {Matsukevich}}, \bibinfo {author}
  {\bibfnamefont {P.}~\bibnamefont {Maunz}}, \ and\ \bibinfo {author}
  {\bibfnamefont {C.}~\bibnamefont {Monroe}},\ }\href@noop {} {\bibfield
  {journal} {\bibinfo  {journal} {Physical Review A}\ }\textbf {\bibinfo
  {volume} {76}},\ \bibinfo {pages} {052314} (\bibinfo {year}
  {2007})}\BibitemShut {NoStop}%
\bibitem [{\citenamefont {Lin}\ \emph {et~al.}(2009)\citenamefont {Lin},
  \citenamefont {Zhu}, \citenamefont {Islam}, \citenamefont {Kim},
  \citenamefont {Chang}, \citenamefont {Korenblit}, \citenamefont {Monroe},\
  and\ \citenamefont {Duan}}]{Lin2009}%
  \BibitemOpen
  \bibfield  {author} {\bibinfo {author} {\bibfnamefont {G.-D.}\ \bibnamefont
  {Lin}}, \bibinfo {author} {\bibfnamefont {S.-L.}\ \bibnamefont {Zhu}},
  \bibinfo {author} {\bibfnamefont {R.}~\bibnamefont {Islam}}, \bibinfo
  {author} {\bibfnamefont {K.}~\bibnamefont {Kim}}, \bibinfo {author}
  {\bibfnamefont {M.-S.}\ \bibnamefont {Chang}}, \bibinfo {author}
  {\bibfnamefont {S.}~\bibnamefont {Korenblit}}, \bibinfo {author}
  {\bibfnamefont {C.}~\bibnamefont {Monroe}}, \ and\ \bibinfo {author}
  {\bibfnamefont {L.-M.}\ \bibnamefont {Duan}},\ }\href@noop {} {\bibfield
  {journal} {\bibinfo  {journal} {EPL (Europhysics Letters)}\ }\textbf
  {\bibinfo {volume} {86}},\ \bibinfo {pages} {60004} (\bibinfo {year}
  {2009})}\BibitemShut {NoStop}%
\end{thebibliography}%
\end{document}